\documentclass{article}
\usepackage{emulateapj}
\usepackage{apjfonts}

\begin{document}

\submitted{To appear in The Astrophysical Journal}

% Keep track of fig and table 
% numbering here because floats are not allowed in preprint format that
% will be used later.

%%% figures
\newcommand{\figrgb}{1}
\newcommand{\figband}{2}
\newcommand{\figcmd}{3}
\newcommand{\figeta}{4}
\newcommand{\figflash}{5}
\newcommand{\figflashzoom}{6}
\newcommand{\figetatemp}{7}
\newcommand{\figetamass}{8}
\newcommand{\figbhevol}{9}
\newcommand{\figredcore}{10}
\newcommand{\figatmos}{11}
\newcommand{\figbandtwo}{12}
\newcommand{\figext}{13}
\newcommand{\figdif}{14}
\newcommand{\fighook}{15}
\newcommand{\figopt}{16}
\newcommand{\figgap}{17}

%%% tables
\newcommand{\tabobs}{1}
\newcommand{\tabcat}{2}

\title{Flash Mixing on the White Dwarf Cooling Curve:
Understanding Hot Horizontal Branch Anomalies in NGC~2808$^1$}

% USE FULL NAME 

\author{Thomas M. Brown$^{2,3}$, 
Allen V. Sweigart,
Thierry Lanz$^4$,
Wayne B. Landsman$^5$, 
Ivan Hubeny$^2$,
}

\affil{Laboratory for Astronomy \& Solar Physics, Code 681, NASA/GSFC,
Greenbelt, MD 20771.\\ tbrown@stsci.edu, 
sweigart@bach.gsfc.nasa.gov, lanz@nova.gsfc.nasa.gov, 
landsman@mpb.gsfc.nasa.gov, hubeny@tlusty.gsfc.nasa.gov}

\begin{abstract}

We present an ultraviolet color-magnitude diagram (CMD) spanning the
hot horizontal branch (HB), blue straggler, and white dwarf
populations of the globular cluster NGC~2808.  These data were
obtained with the far-UV and near-UV cameras on the Space Telescope
Imaging Spectrograph (STIS).  Although previous optical CMDs of
NGC~2808 show a high temperature gap within the hot HB population, no
such gap is evident in our UV CMD.  Instead, we find a population of
hot subluminous HB stars, an anomaly only previously reported for the
globular cluster $\omega$ Cen.  Our theoretical modeling indicates
that the location of these subluminous stars in the UV CMD, as well as
the high temperature gap along the HB in optical CMDs, can be
explained if these stars underwent a late helium-core flash while
descending the white dwarf cooling curve.  We show that the convection
zone produced by such a late helium flash will penetrate into the
hydrogen envelope, thereby mixing hydrogen into the hot helium-burning
interior, where it is rapidly consumed.  This phenomenon is analogous
to the ``born again'' scenario for producing hydrogen-deficient stars
following a late helium-shell flash.  The flash mixing of the envelope
greatly enhances the envelope helium and carbon abundances, and leads,
in turn, to a discontinuous increase in the HB effective temperatures
at the transition between canonical and flash-mixed stars.  We argue
that the hot HB gap is associated with this theoretically predicted
dichotomy in the HB properties.  Moreover, the changes in the emergent
spectral energy distribution caused by these abundance changes are
primarily responsible for explaining the hot subluminous HB stars.
Although further evidence is needed to confirm that a late helium-core
flash can account for the subluminous HB stars and the hot HB gap, we
demonstrate that an understanding of these stars requires the use of
appropriate theoretical models for their evolution, atmospheres, and
spectra.

\end{abstract}


\section{INTRODUCTION} \label{secintro}

The horizontal branch (HB) represents the core helium-burning phase of
stellar evolution for low-mass stars.  This late phase of stellar
evolution is named for its theoretical locus in the HR diagram, which
spans a much wider range in temperature than in bolometric luminosity.
The observed temperature distribution of HB stars tends to become
redder at increasing metallicity, and thus metallicity is generally
regarded as the ``first parameter'' of HB morphology.  However, the
existence of clusters with similar metallicities but very distinct HB
morphologies has led to the ``second parameter'' debate.  Indeed, Rich
et al.\ (1997\markcite{R97}) have recently shown that even metal-rich
globular clusters can have extended blue HBs.  It is well-known that
multiple parameters can potentially act as second parameters on the HB
(e.g., age, mass loss, helium abundance, rotation, cluster dynamics;
Fusi Pecci \& Bellazzini 1997\markcite{FB97}).  Although all of these
parameters can theoretically govern HB morphology, it remains to be
seen which of them actually plays a dominant role. \\ 

{\small 
$^1$Based on observations with the NASA/ESA Hubble Space
Telescope obtained at the Space Telescope Science Institute, which is
operated by AURA, Inc., under NASA contract NAS~5-26555.

$^2$NOAO

$^3$Now at the Space Telescope Science Institute, 3700 San Martin Drive,
Baltimore, MD 21218. 

$^4$University of Maryland

$^5$Raytheon ITSS
} \\

Besides the second parameter effect, there are many other intriguing
peculiarities in the HB morphology that still defy explanation.  Some
globular clusters possess bimodal HBs with substantial populations of
both ``blue'' HB (BHB) stars hotter than the instability strip, and
``red'' HB (RHB) stars cooler than the instability strip, but with
very few stars in between.  In some clusters one finds a long blue HB
trail, often punctuated by gaps, that extends to the ``extreme'' HB
(EHB) stars, which are defined in this paper as HB stars with
effective temperatures $\rm T_{eff} \gtrsim 20,000$~K.  These EHB
stars correspond to the subdwarf B (sdB) stars studied in the Galactic
field population (e.g., Saffer et al.\ 1994\markcite{SBK94}).  Another
peculiarity has been reported by D'Cruz et al.\ (2000\markcite{D00}),
who found a population of subluminous HB stars in $\omega$~Cen.

The HB is not simply a sequence in $\rm T_{eff}$ -- it is also a
sequence in envelope mass ($M_{env}$).  Stars with large $M_{env}$
occupy the RHB, while stars with small $M_{env}$ occupy the EHB.  In
theory, the range in $M_{env}$ can result from a range in
red-giant-branch (RGB) mass loss.  However, it is difficult to see how
one mechanism alone can produce the entire range of HB stars.  A
low-mass star would have to lose several tenths of a solar mass on the
RGB, while retaining an envelope mass of just a few hundredths of a
solar mass, in order to arrive on the hot end of the zero-age HB
(ZAHB) via the same process that produces stars on the cool end of the
ZAHB.  Furthermore, the gaps on the HB would then imply ``forbidden''
values of mass loss and/or envelope mass.  Historically, populating
the entire HB via one mechanism has implied an unpalatable fine-tuning
of the mass-loss process.  Instead, there may be distinct physical
processes populating different temperature ranges on the HB (see,
e.g., Ferraro et al.\ 1998\markcite{F98}).  Peculiarities in HB
morphology, such as gaps and subluminous stars, may be important clues
to these formation mechanisms.

In this paper, we analyze new ultraviolet images of NGC~2808, a
globular cluster that has been well-studied because of its unusual HB
morphology.  NGC~2808 was one of the first clusters known to have a
bimodal HB, with a large gap between the BHB and RHB stars (Harris
1974\markcite{H74}) and only 2 known RR Lyrae (Clement \& Hazen
1989\markcite{CH89}).  Rood et al.\ (1993\markcite{R93}) have pointed
out that NGC~2808 may even be an example of the second parameter
effect operating within a single cluster.  Although NGC~2808 is of
intermediate metallicity ([Fe/H]~$ =-1.36$; Walker
1999\markcite{W99}), it possesses one of the longest blue HB tails of
any globular cluster.  Moreover, there are two striking gaps within
this blue tail: one between the EHB and BHB, and another within the
EHB itself (Sosin et al.\ 1997\markcite{S97}; Walker
1999\markcite{W99}; Bedin et al.\ 2000\markcite{BPZ00}); the colors of
the gaps imply effective temperatures of 17,000~K and 25,000~K,
respectively, if the color-temperature transformation employs
synthetic spectra at the cluster metallicity.  Although unusual, HB
gaps are not unique to NGC 2808; gaps have also been found in many
other globular clusters (e.g., M13, M80, and NGC~6273; Ferraro et al.\
1998\markcite{F98}; Piotto et al.\ 1999\markcite{P99}), while Whitney
et al.\ (1994\markcite{W94}, 1998\markcite{WRO98}) and D'Cruz et al.\
(2000\markcite{D00}) have reported both gaps and subluminous stars in
the HB population of $\omega$ Cen.  Our UV images of NGC~2808 confirm
the presence of a pronounced gap between the EHB and BHB, but not the
gap seen within the EHB in optical color magnitude diagrams (CMDs).
Instead, we find a substantial population of subluminous EHB stars.
In this paper, we will present a new theoretical scenario for
understanding these peculiarities of the EHB population in NGC~2808,
as well as $\omega$~Cen.

D'Cruz et al.\ (2000\markcite{D00}) and Whitney et al.\
(1998\markcite{WRO98}) claimed that the hot subluminous stars on the
HB of $\omega$ Cen were probably a population of ``blue-hook'' HB
stars.  These are stars that experience a late helium-core flash.
Normally, stars ignite He burning at the tip of the RGB, but if they
undergo very high mass-loss, the stars will leave the RGB and ignite
He burning while descending the white dwarf (WD) cooling curve
(Castellani \& Castellani 1993\markcite{CC93}).  D'Cruz et al.\
(1996\markcite{D96}) suggested that such stars provide another avenue
for populating the hot end of the HB, thus helping to alleviate the
fine-tuning problem.  In their scenario, the He-burning stars that
arise from such a late flash should form a ``blue-hook'' appended to
the hot end of the canonical EHB, with slightly lower core masses (by
$\sim$0.01 $M_{\odot}$) than those stars that arrive on the ZAHB
directly from the RGB.  D'Cruz et al.\ (2000\markcite{D00}) claimed
that blue-hook stars should lie up to 0.1~mag below the EHB, but the
subluminous HB stars in their CMD of $\omega$ Cen lie up to
$\sim$0.7~mag below the canonical EHB and, in addition, span a wide
range in color, demonstrating that this idea needs further
exploration.

Sweigart (1997\markcite{S97}) has shown that when stars undergo a late
helium-core flash on the WD cooling curve, flash mixing of the
hydrogen envelope with the helium core will greatly enhance the
envelope helium and carbon abundances.  At the peak of the flash, the
He-burning luminosity ($L_{He}$) reaches 10$^{10}$ $L_{\odot}$.
Normally, the flash convection produced by such a high burning rate
does not penetrate the hydrogen shell; however, when stars ignite He
on the WD cooling curve, they do so in the presence of a much weaker
hydrogen-burning shell.  The lower entropy barrier associated with
this weaker hydrogen-burning shell (Iben 1976\markcite{I76}) allows
the flash convection to penetrate the hydrogen envelope, thereby
mixing hydrogen into the hot He-burning interior, where it is rapidly
consumed.  The result is a hydrogen-deficient He-burning star with
enhanced helium and carbon in the envelope.  A similar process has
been proposed to explain the extremely hydrogen-deficient R Coronae
Borealis (R~CrB) stars (also known as ``born again'' stars), whereby
the envelope hydrogen is mixed inward during a very late helium-shell
flash (e.g., Renzini 1990\markcite{R90}).  This flash mixing might
also explain the enhanced helium abundances seen in some field sdO
stars (Lemke et al.\ 1997\markcite{L97}) and a small fraction of the
field sdB stars (e.g., Moehler et al.\ 1990\markcite{M90}).  Because
flash mixing produces envelope abundances that are very distinct from
those in canonical EHB stars, the stellar atmospheres used for the
models of blue-hook stars should include these abundance changes when
predicting their observed colors and luminosities.

In this paper, we will examine new observations of NGC~2808, taken
with the Space Telescope Imaging Spectrograph (STIS) on board the
Hubble Space Telescope (HST).  In \S\ref{secobs}, we describe the
observations, data reduction, and photometry.  We then present the UV
CMD in \S\ref{secuvcmd}, revealing a population of stars below the
canonical ZAHB, a phenomenon previously seen only in $\omega$ Cen.  In
\S\ref{secevol}, we present the stellar evolutionary tracks we use to
investigate the HB morphology, with special attention to models with
flash mixing on the WD cooling curve.  The atmospheres of flash-mixed
stars are expected to have hydrogen-poor, non-solar scaled abundances;
our adopted model atmospheres with these abundances are discussed in
\S\ref{secspec}.  In \S\ref{secsubhb}, we first rule out several
alternative scenarios for understanding the hot sub-ZAHB stars, and
then conclude that flash mixing provides the most likely explanation.
We then demonstrate, in \S\ref{secgap}, that the dichotomy between the
flash-mixed stars and the canonical ZAHB can account for the EHB gap
seen in optical CMDs of NGC~2808.  We consider some implications of
the flash-mixing scenario in \S\ref{secimp}.  We then review our
conclusions in \S\ref{secsum}.

\section{OBSERVATIONS AND DATA REDUCTION} \label{secobs}

\subsection{Imaging} \label{secimg}

Because NGC~2808 provides a well-sampled field of hot stars, the
cluster was observed with STIS in order to provide a geometric
distortion correction for the STIS camera modes.  Although the
observation plan was driven by the calibration goals, such calibration
data, which are made immediately available to the public, can also be
of scientific interest.  Images were obtained in the far-UV crystal
quartz (FUV/F25QTZ; pivot wavelength 1600~\AA), near-UV 2700~\AA\
continuum (NUV/F25CN270), and clear CCD (CCD/50CCD; pivot wavelength
5850~\AA) modes, employing a cross-shaped dither pattern.  The UV
images were taken in the cluster center, but most of the CCD images
were offset from the cluster center, to image a less crowded field.
The CCD images obtained in the center are too crowded for useful
photometry; we do not include them in our analysis, and employ them
only to provide a third color for a false-color image of the cluster
center.  For the UV images that are the focus of this paper, the
resulting sky coverage is almost three times larger than a single STIS
exposure, but the depth of this coverage is very nonuniform, due to
the overlapping coverage between individual exposures.  Nonetheless,
hot stars on and near the horizontal branch are well-detected in
single exposures, so this varying depth only significantly affects the
detection and measurement of stars that are on the RHB or much fainter
than the ZAHB.  The observational parameters for the UV images are
summarized in Table~\tabobs.  The exposure times reflect the total
amount of time observing in each mode (not the exposure depth at any
one point in the image).  A subsection of the STIS images is shown in
Figure~\figrgb.  The UV bandpasses are shown in Figure~\figband. \\ 

\smallskip

%\begin{table}
%\caption{UV Imaging}
%\hskip 0.5in

\noindent
\parbox{3.25in}{
{\sc Table \tabobs:} UV Imaging 

\begin{tabular}{lcccc}
\tableline
STIS         & Exp.  & Observation      & coverage         \\  
band         & (sec)     & date (2000)      & ($\sq\arcsec$)  \\ 
\tableline			  		     		  
FUV/F25QTZ   & 8916      & Jan 18 \& Feb 16 & 1744             \\
NUV/F25CN270 & 8906      & Jan 19 \& Feb 20 & 1752             \\
\tableline
\end{tabular}
%\end{table}
}
\medskip

The UV detectors employed by STIS are multi-anode microchannel arrays
(MAMAs) with very little red leak.  The MAMAs are also photon counters
that register less than one count per incident cosmic ray; thus,
cosmic ray rejection is not required for the UV imaging.  For the CCD
imaging (where an incident cosmic ray causes a massive many-count
``hit''), we employed standard cosmic-ray rejection.  A full
description of the instrument and its capabilities can be found in
Woodgate et al.\ (1998\markcite{W98}) and Kimble et al.\
(1998\markcite{K98}).

As explained above, the individual exposures were obtained in a
cross-shaped dither pattern spanning the center of the cluster.  Thus,
we needed to rotate and register them before co-adding.  The plate
scale for the far-UV and near-UV modes differs slightly, but both are
within a few percent of $0.025\arcsec$ pix$^{-1}$.  The plate scale
of the CCD images is $0.05\arcsec$ pix$^{-1}$.  Using the DRIZZLE
package (Fruchter \& Hook 1998\markcite{FH98}) in IRAF, we drizzled
the individual exposures in each mode to the same plate scale (exactly
$0.025\arcsec$ pix$^{-1}$), including a correction for the geometric
distortion.  We then cross-correlated those images to solve for
relative rotation and offsets, and then re-drizzled the original
$1024\times1024$ pixel exposures to create a summed $2048\times 2048$
pixel image for each band.  The drizzling included masks for occulted
corners and edges of the detector, plus hot and defective pixels.

Spectroscopic monitoring of the STIS UV modes has shown a small
($\lesssim 2$\%) variation in the near-UV sensitivity over the course
of almost three years, with no net loss of sensitivity.  However, the
far-UV spectroscopic sensitivity has shown a small net decline (Bohlin
1999\markcite{B99}); this decline is somewhat wavelength dependent,
ranging from 0.8\%--2.8\% per year across the entire far-UV wavelength
range.  The near-UV sensitivity variations are small enough to be
ignored, but to minimize systematic errors, we derived a new far-UV
imaging throughput curve for the date of our far-UV exposures.  This
new throughput curve is the product of the nominal throughput curve
and the spectroscopic sensitivity drop as of 1 Feb 2000.  This date is
midway between the two sets of far-UV exposures, taken a month apart
(see Table~\tabobs); the change in the sensitivity over that month is
insignificant.  A flat $F_\lambda$ spectrum produces 7\% fewer counts
with the new far-UV sensitivity curve, compared to the nominal curve.
Our revised far-UV imaging bandpass is shown in Figure~\figband, and
it will be used to translate the theoretical stellar models to the
observational plane.

\subsection{Spectroscopy} 

The brightest star in these images was also observed spectroscopically
with STIS on 20 April 1999.  This star is located $\sim 18\arcsec$
from the cluster center; in Figure~\figrgb, it is the bright blue star
on the upper right.  The observations were taken as part of a
spectroscopic followup program (HST Guest Observer program 7436) of
post-asymptotic giant branch (post-AGB) candidates discovered on
globular cluster images obtained with the Ultraviolet Imaging
Telescope (UIT; Landsman et al.\ 2001\markcite{L01}).  The
spectroscopy was obtained through the $52\arcsec \times 0.2\arcsec$
slit in the G140L, G230L, and G430L modes, providing wavelength
coverage from 1150~\AA\ to 5700~\AA.  

We reduced the spectroscopic data with the CALSTIS package of the
STIS Instrument Definition Team (Lindler 1999\markcite{L99}).  The
star is sufficiently bright and isolated to allow the use of the
standard extraction slit for the source (11 pixels).  The reduced
spectrum is shown in Figure~\figband.  We use this spectrum to tie our
photometry to an absolute flux scale, as described in the next
section.

\subsection{Photometry} \label{secphot}

Because both colors and luminosities are needed for the CMD, we are
only interested in photometry for stars that are well-detected in both
the far-UV and near-UV bands.  To create a coordinate list of stars in
the STIS images, we centroided on each star detected by eye in the
far-UV image.  The far-UV image is sparsely populated, and it has
nonuniform coverage with varying signal-to-noise ratio, so this
yielded much better results than an automated detection algorithm.

Once we had an object list for the cluster, we used the IRAF routine
PHOT in the DIGIPHOTX/DAOPHOTX package to perform small-aperture
photometry.  The object aperture radius was 4 pixels, and the sky
annulus spanned radii of 4--10 pixels.  The sky annulus thus included
a small predictable amount of source flux from each target star, but
more accurately reflected the local background from the wings of
neighboring stars.  Because STIS is optimized for spectroscopy instead
of imaging, the point spread function (PSF) varies as a function of
horizontal position in a given exposure; our final images are the sum
of many dithered exposures, which tends to average out the variations
in the PSF, but use of an object aperture radius smaller than 4 pixels
would increase the photometric scatter.  To put the NGC~2808
photometry on an absolute scale, we used our spectrum of the
UV-bright post-AGB star in the cluster (see Figure~\figband) to
predict the imaging count rate for this star, because the
spectroscopic sensitivity is far easier to calibrate than imaging
sensitivity (through comparison with WD spectra).  Comparison to the
measured imaging count rate gave an aperture correction of 1.83 for
the FUV/F25QTZ photometry and 1.44 for the NUV/F25CN270 photometry.
Note that these aperture corrections account for both the encircled
energy within the source aperture and also the subtraction of source
light included in the sky annulus.  Based upon an analysis of the STIS
point spread function (PSF; Robinson 1997), the near-UV aperture
correction is 2\% smaller than expected, while the far-UV aperture
correction is 6\% larger than expected.  The larger far-UV discrepancy
reflects telescope breathing and systematic uncertainties in the
imaging calibration at the $\lesssim$0.1~mag level.  We felt that
normalization to this stellar spectrum would provide a more accurate
photometric scale.  In any case, the differences between our aperture
correction and the nominal corrections for the STIS PSF are small.

Magnitudes for our photometry are in the STMAG system:\\
\[m = {\rm -2.5 \times log_{10}} f_{\lambda} -21.10\]
\[f_{\lambda} ={\rm counts \times PHOTFLAM / EXPTIME}\] \\ \\

%\begin{figure}
%\plotone{fig1.eps}
%\caption{
\hskip 0.25in
\parbox{6.5in}{\epsfxsize=6.5in \epsfbox{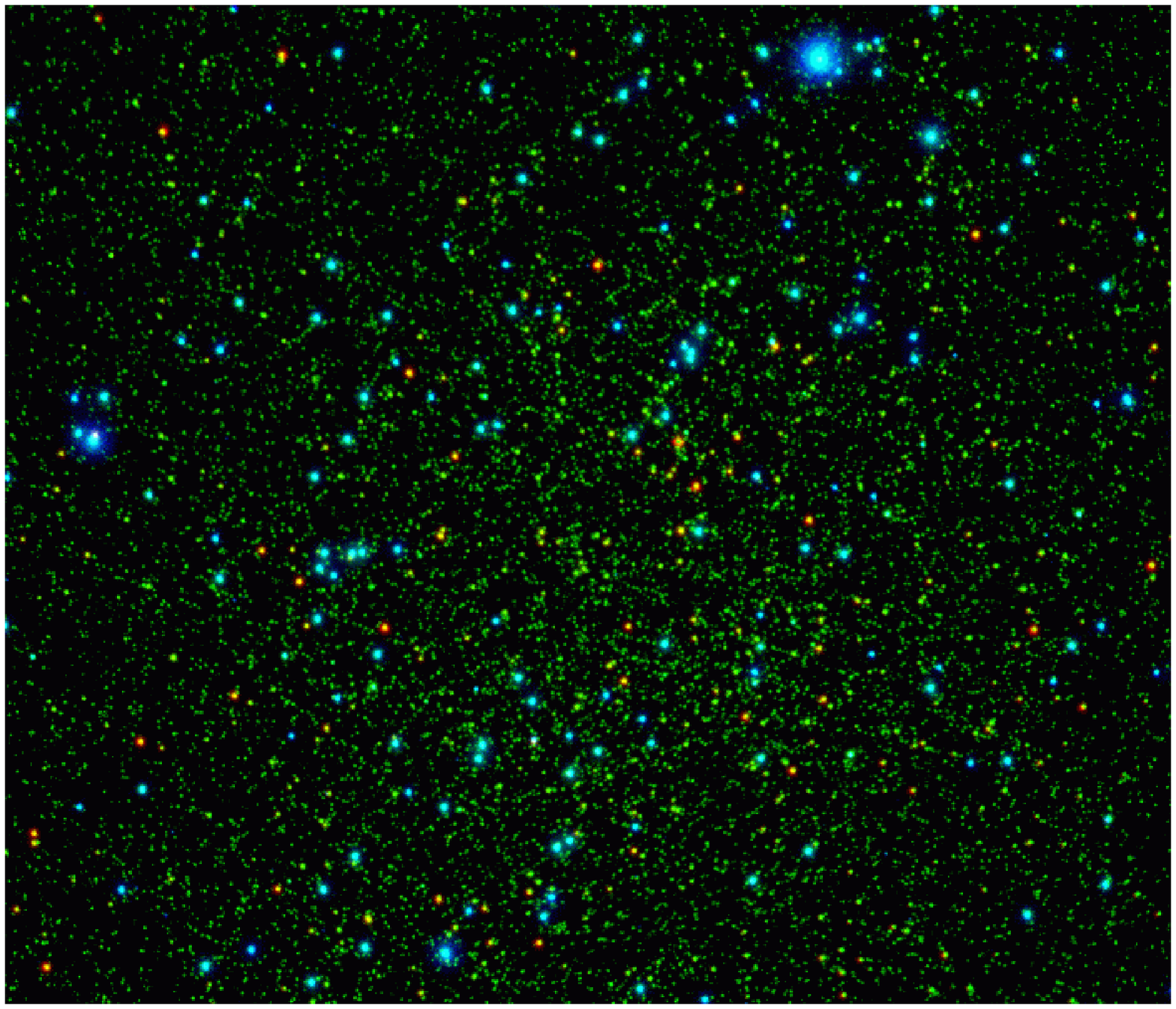}} \\

\hskip 0.25in
\parbox{6.5in}{\small {\sc Fig.~\figrgb--}
STIS observations of the evolved stellar populations in the
center of NGC~2808.  This false-color image is the result of assigning
the STIS far-UV image to the blue channel, the near-UV image to the
green channel, and the optical CCD image to the red channel.  The
image has been cropped to $35\arcsec \times 30\arcsec$ for display
purposes.  The population is well-resolved at the center of this dense
cluster.  We have also obtained UV-to-optical spectroscopy of the
bright blue post-AGB star near the top of the image (see Figure~\figband).}
%\end{figure} 

\medskip

\noindent
where EXPTIME is the exposure time, and PHOTFLAM is 
$1.11 \times 10^{-16}$ erg s$^{-1}$ cm$^2$ \AA$^{-1}$ / (cts s$^{-1}$)
for the FUV/F25QTZ filter, and 
$3.29 \times 10^{-17}$ erg s$^{-1}$ cm$^2$ \AA$^{-1}$ / (cts s$^{-1}$)
for the NUV/F25CN270 filter.  Note that this PHOTFLAM for the far-UV
filter takes into account the 7\% reduction in far-UV sensitivity.

Our photometric catalog contains 295 stars with $m_{FUV} \leq 22$~mag,
$m_{NUV} \leq 22$~mag, and photometric errors $\leq$ 0.2 mag; the
catalog is available in Table~\tabcat.  The far-UV
image is very sparsely populated, with fewer than one star per 1000
resolution elements to a depth of $m_{FUV} = 22$~mag.  The near-UV
image is fairly crowded, but not to an extent that affects our
photometry.  If automated object detection and photometry are done on
the near-UV image (instead of using the position catalog derived from
the far-UV), we find that there are fewer than one star per 30
resolution elements to a depth of $m_{NUV} = 22$~mag, fewer than one
star per 55 resolution elements to a depth of $m_{NUV} = 21$~mag, and
fewer than one star per 300 resolution elements to a depth of $m_{NUV}
= 20$~mag.\\

\vskip 6.5in

\section{Ultraviolet Color Magnitude Diagram} \label{secuvcmd}

We present the STIS photometry as a CMD in Figure~\figcmd.  The data
are shown as error bars, and the theoretical fiducials are shown as
labeled grey curves.  The exposure depth is variable across the image,
so the photometric errors do not uniformly increase toward fainter
magnitudes, but all stars on or near the HB are well-detected.  The
EHB and BHB stars fall into two distinct clumps, with 75 stars and 128
stars, respectively.  White dwarf and blue straggler (BS) stars are also
present.  Below, we review the cluster parameters, evolutionary
models, and synthetic spectra that drive the location of the
theoretical fiducials.  We then point out peculiarities in our CMD,
and in the subsequent sections explore the theoretical interpretation
in more detail.

\subsection{Cluster Parameters}

The metallicity, distance modulus, and foreground extinction are
somewhat uncertain for NGC~2808, and have been examined extensively in
the literature (e.g., Walker 1999\markcite{W99} and references
therein).  Walker (1999\markcite{W99}) makes a compelling case that
[Fe/H]$=-1.36 \pm 0.05$ on the Zinn \& West (1984\markcite{ZW84})
metallicity scale (see also Rutledge, Hesser, \& Stetson
1997\markcite{R97}; Ferraro et al.\ 1999\markcite{F99}), and that is
the value we will adopt here.  Recently, Bedin et al.\
(2000\markcite{BPZ00}) calculated the extinction and distance modulus
for this cluster using several different methods and metallicity
assumptions.  When assuming the Zinn \& West (1984\markcite{ZW84})
metallicity scale, they found that the distribution of RGB stars
implies $E(B-V)=0.18 \pm 0.01$ and $(m-M)_V=15.60 \pm 0.10$.  These
are the values we will adopt here, which give good agreement between
the theoretical and observed BHB locus.  We apply this foreground
reddening to all of the synthetic spectra using the Cardelli, Clayton,
\& Mathis (1989\markcite{CCM89}) parameterization.  Assuming $A_V =
3.1 \times E(B-V) = 0.56$~mag, this gives $(m-M)_o = 15.04$ and a
distance of 10.2 kpc. \\

%\begin{figure}
%\plotfiddle{fig2.eps}{3.0in}{0}{100}{100}{-250}{0}
%\caption{ 
\parbox{3.25in}{\epsfxsize=3.25in \epsfbox{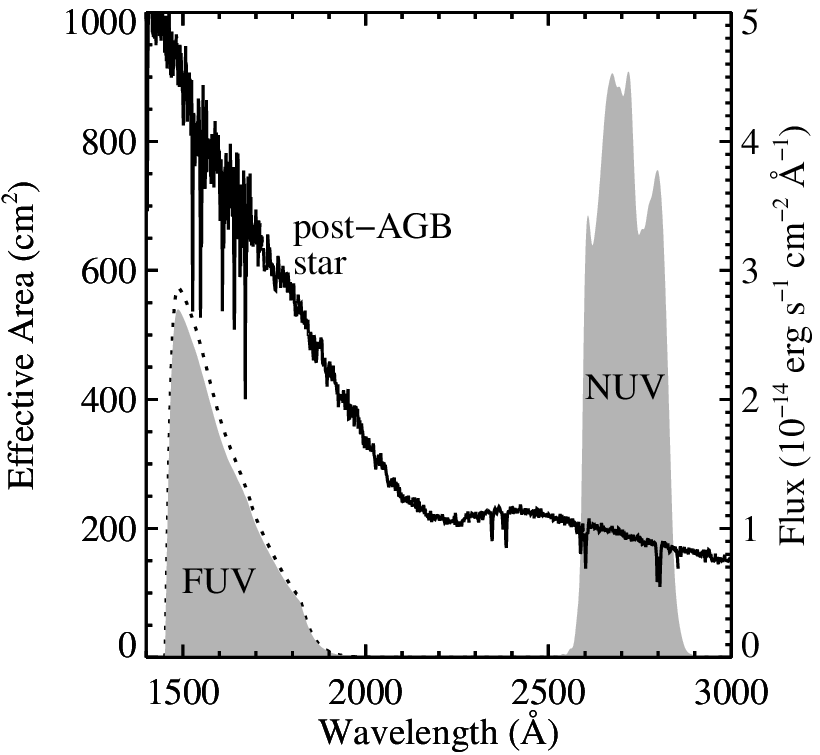}} \\

\parbox{3.25in}{\small {\sc Fig.~\figband--}
STIS far-UV and near-UV bandpasses (grey filled areas; left-hand 
scale) used for the NGC~2808 imaging.  The dotted line shows the
nominal far-UV bandpass at the start of the STIS mission; at the time
of the NGC~2808 imaging, the sensitivity was degraded by 7\%.  We also
show a portion of the spectrum for the post-AGB star observed in
NGC~2808 (right-hand scale), used to ensure that our aperture
photometry was on an absolute flux scale. }
%\end{figure} 

\medskip

\subsection{White Dwarfs}

White dwarfs are not the focus of this paper, but their possible
presence in Figure~\figcmd\ will be of interest to some readers.  To
distinguish stars that may lie on the WD cooling curve, we employed
the 0.5 and 0.6~$M_{\odot}$ C/O-core sequences of Wood
(1995\markcite{W95}), representing DA WD stars (the actual WD stars in
NGC~2808 are likely to have a mass near 0.55~$M_{\odot}$).  To
translate the WD sequence to the observational plane, we used the
TLUSTY model atmosphere code (Hubeny \& Lanz 1995\markcite{HL95}) to
calculate a sequence of non-LTE, pure hydrogen model atmospheres, and
then used the SYNSPEC spectra synthesis code (Hubeny, Lanz \& Jeffery
1994\markcite{HLJ94}) to calculate spectra.  This was done for
temperatures hotter than 20,000~K in the 0.5 and 0.6~$M_{\odot}$
cooling curve sequences (at cooler temperatures, these sequences fall
below the detection limits in our CMD). The synthetic spectra have no
lines other than those of hydrogen.

If high RGB mass-loss does indeed drive the hot HB morphology in
NGC~2808, the cluster might also contain a population of helium-core
white dwarfs that never core flash.  A helium-core WD would have a
larger radius than a C/O-core WD at the same $\rm T_{eff}$ (e.g.,
Panei, Althaus, \& Benvenuto 2000\markcite{P00}).  Unfortunately,
because the WD cooling tracks in the STIS CMD (Figure~\figcmd) are
nearly vertical, the helium WD cooling tracks of Panei et al.\
(2000\markcite{P00}) are indistinguishable from the C/O-core cooling
tracks already plotted.  Note that the presence of helium WDs in
NGC~2808 might provide a counterexample to the usual assumption that
helium WDs can only be formed as the products of binary evolution.

\subsection{Blue Stragglers}

The BS sequence in NGC~2808 has been seen clearly in
previous optical CMDs, and the brightest blue stragglers are also
present in the STIS data.  The BS sequence should roughly
follow the zero-age main sequence (ZAMS) up to twice the turnoff mass,
if BS stars are the result of mergers between two MS stars.
The MS turnoff mass in NGC~2808 is approximately 0.85~$M_{\odot}$,
and well below the detection limits in our UV images, but stars near
twice the turnoff mass are easily detectable in our STIS images.  To
highlight the BS sequence in these clusters, we used an isochrone from
Bertelli et al.\ (1994\markcite{B94}), at [Fe/H]=$-1.3$ and an age of
4~Myr, truncating the isochrone at 1.7~$M_{\odot}$.  The faintest part
of the ZAMS shown in the CMD corresponds to 1.2~$M_{\odot}$.  Assuming
the nominal cluster abundances, we used the Kurucz
(1993\markcite{K93}) grid of LTE synthetic spectra to translate the
ZAMS to the observational plane, interpolating in effective
temperature and metallicity from the grid points that most closely
matched each point along the ZAMS.  Note that depending upon the
formation mechanism, blue straggler stars might have unusual surface
abundances (e.g., Bailyn 1992\markcite{B92}), which would produce
scatter away from the theoretical ZAMS.

\subsection{Horizontal Branch Stars} \label{seccmdhb}

Of primary concern in this paper is the distribution of HB stars in
our UV CMD.  The theoretical HB fiducials in Figure~\figcmd\ come from
our own calculations, which are discussed extensively in
\S\ref{secevol} and \S\ref{secspec}.  These fiducials assume canonical
evolution from the main sequence to the core He-burning phase, and
were translated to the observational plane with synthetic spectra at
the cluster metallicity.  The lower bound of the canonical HB locus is
the zero-age HB.  The upper bound of the canonical HB locus marks the
point where the star has completed 99\% of its core He-burning
lifetime; beyond this point, the evolution proceeds relatively
rapidly, so that few stars would be expected.

In the UV, the HB morphology of NGC~2808 shares several features with
CMDs at longer wavelengths, but also shows significant differences.
The gap between the EHB and BHB, at $m_{FUV}-m_{NUV} \sim -1$~mag, is
well detected, as is the gap between the BHB and RHB, at
$m_{FUV}-m_{NUV} > 0$~mag.  In fact, because the temperature scale is
greatly stretched as one goes to cooler $\rm T_{eff}$, the BHB-RHB gap
is exaggerated, and the RHB stars are not detected at these
wavelengths.  However, the gap within the EHB (Sosin et al.\
1997\markcite{SDD97}; Walker 1999\markcite{W99}; Bedin et al.\
2000\markcite{BPZ00}) is not present in our UV CMD.  Instead, the EHB
in the UV has a much larger luminosity width than the theoretical EHB,
with many stars falling well below the theoretical EHB.  Out of a
total EHB population of 75 stars in Figure~\figcmd, 29 lie within the
canonical EHB locus, while 46 are subluminous.

Given the large number of EHB stars, one expects to find a population
of post-HB progeny.  There are, in fact, 6 hot stars in our CMD that
lie at luminosities far above the HB.  The brightest is a post-AGB
star, for which we have obtained UV-optical 

%\begin{figure}
%\plotfiddle{fig3.eps}{6.0in}{0}{100}{100}{-250}{45}
%\vskip -0.9in
%\caption{\small 

\hskip 0.25in
\parbox{6.5in}{\epsfxsize=6.5in \epsfbox{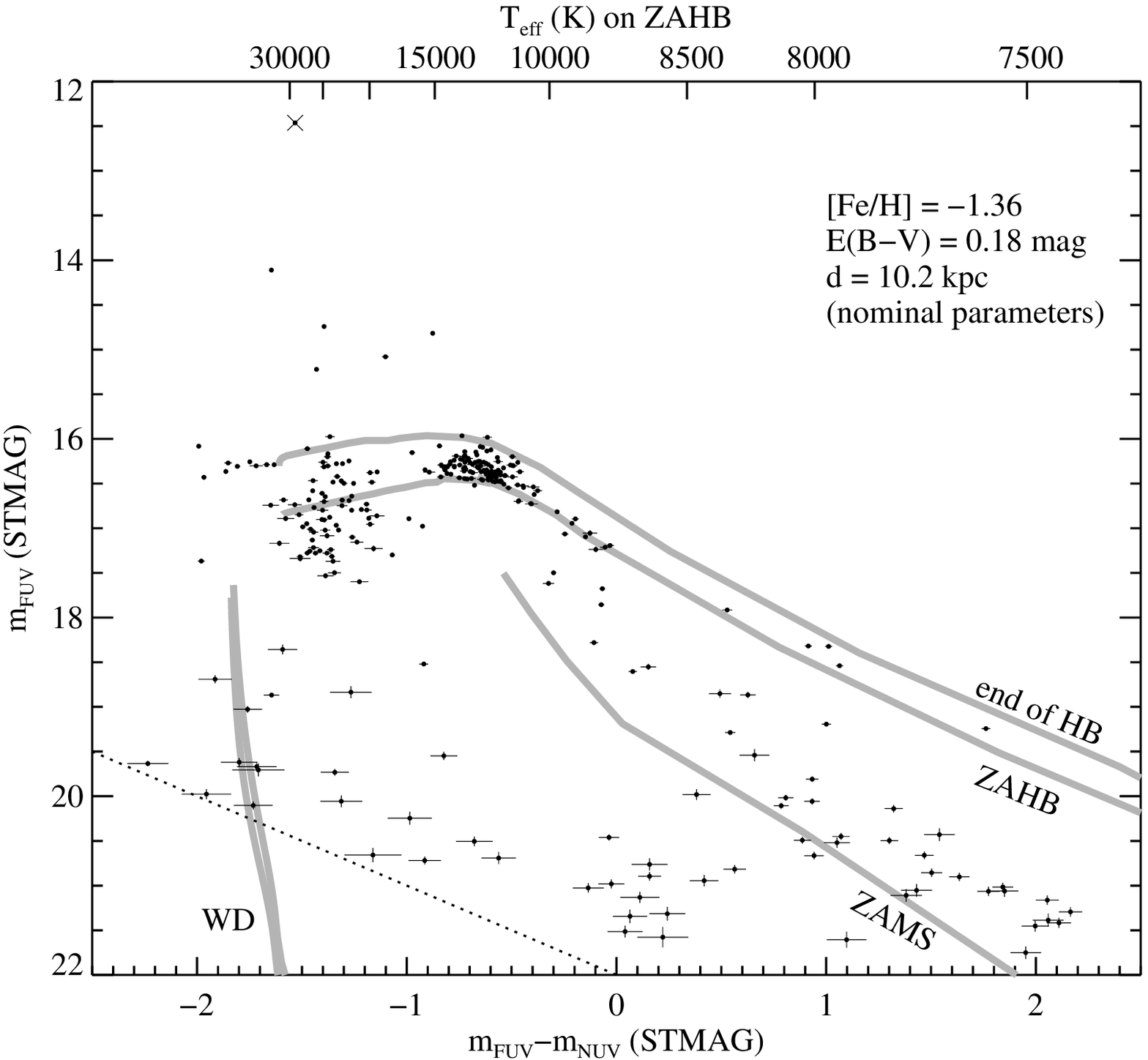}} \\ 

\hskip 0.25in
\parbox{6.5in}{\small {\sc Fig.~\figcmd--}
The STIS UV CMD for NGC~2808.  The different stages of
stellar evolution are traced by grey curves that were transformed to
the observational plane assuming the nominal cluster parameters. The
exposure depth is variable across the image, so the photometric errors
do not uniformly increase in size toward fainter magnitudes, but all
stars on or near the HB are well-detected.  The lower bound of the HB
locus is the zero-age HB, and the upper bound is the point where the
star has completed 99\% of its core He-burning lifetime.  One post-AGB
star appears in the CMD (X), and was also observed spectroscopically
by STIS (see Figure~\figband).  The locus of BHB stars falls within
the theoretical boundaries, but the EHB stars spread well below the
ZAHB.  Five stars that lie $\sim$1~mag above the EHB fall in the
expected location for post-HB stars.  The BS sequence falls near an
extension of the ZAMS (labeled; Bertelli et al.\
1994\protect\markcite{B94}), truncated at 1.7~$M_{\odot}$.  A faint WD
sequence is also detected (labeled; Wood 1995\protect\markcite{W95}).
The dotted line represents the catalog limit at
$m_{FUV}=m_{NUV}=22$~mag.}
%\end{figure} 

\medskip

\noindent
spectroscopy (see
Figure~\figband).  The other five stars are post-EHB stars; their
position in the CMD agrees well with the slowest phase of
AGB-Manqu$\acute{\rm e}$ evolution, which has a lifetime that is
$\sim$5 times shorter than the HB lifetime.  Thus, there should be
$\sim$5 EHB stars for every post-EHB star, in good agreement with the
number of stars within the canonical EHB locus.  Note that a few of
these 29 stars might be the post-HB progeny of the subluminous EHB
stars, instead of normal EHB stars.  However, the number of post-EHB
stars is considerably smaller than expected if one assumes they are
the progeny of the entire EHB population (75 stars).  Interestingly,
in NGC~6752, Landsman et al.\ (1996) found only four post-EHB stars
descended from a population of 63 apparently normal EHB stars.

There is also a string of 9 stars at $16 < m_{FUV} < 16.5$~mag and $-2
< m_{FUV}-m_{NUV} < -1.6$~mag that lie hotter and brighter \\

\vskip 7.15in
\noindent
than the
hot end of the HB.  These stars are too hot and faint to be normal
AGB-Manqu$\acute{\rm e}$ stars.  Indeed, the bluest of these 9 stars
have colors that would imply temperatures well over 100,000~K,
depending upon the spectra used for the color transformation.  Strong
emission lines (e.g., \ion{C}{4}$\lambda 1550$~\AA) could provide
enough flux in the narrow far-UV bandpass to significantly shift the
color towards the blue, but we have no reason to expect such emission
in hot HB or post-HB stars.  The stars are also too numerous to be hot
WDs, else we would see far more stars along the WD cooling curve.
Cataclysmic variables (CVs) with strong \ion{C}{4} and \ion{He}{2}
emission in the far-UV have been found in the globular cluster 47~Tuc
(Knigge et al.\ 2000\markcite{K00}), but CVs are much too faint and
cool to provide a viable explanation.

The most intriguing possibility is that these 9 hot stars may be the
AGB-Manqu$\acute{\rm e}$ progeny of the subluminous EHB stars.
Indeed, theoretical lifetimes predict that there should be $\sim$9
such post-HB stars for the 46 subluminous EHB stars.  We will briefly
return to this possibility in \S\ref{secobstest}, after we discuss the
origin of the subluminous EHB stars.

Although the post-HB stars are interesting for further study, we are
concerned in the present work with two specific peculiarities of the
HB in NGC~2808: the subluminous EHB stars in our UV CMD, and the EHB
gap in previous optical CMDs.  These features are not predicted by
theoretical fiducials based upon canonical stellar evolution theory,
model atmospheres, and synthetic spectra.  Thus, in \S\ref{secevol},
we explore in greater detail both the canonical HB evolution and the
new evolutionary paths associated with high RGB mass loss (the
flash-mixing scenario).  This will drive the calculation of model
atmospheres and synthetic spectra with abundance enhancements and
departures from LTE, discussed in \S\ref{secspec}.  In the subsequent
sections, we will return to the observations and consider alternative
explanations for these unusual features of the HB morphology in
NGC~2808.

\section{HORIZONTAL BRANCH EVOLUTION} \label{secevol}

\subsection{Sequence Parameters} \label{secseqpar}

A major goal of this paper is to determine if the ``flash-mixing''
scenario outlined in \S\ref{secintro} can account for the hot
subluminous stars and the EHB gap in NGC~2808.  To explore this
possibility, we have constructed a detailed grid of evolutionary
sequences that follow the evolution of appropriate low-mass stellar
models from the ZAMS through the HB phase and, in many cases, through
the subsequent post-HB phases to the WD cooling curve.  Most
importantly, these calculations follow the evolution continuously
through the helium flash to the ZAHB.  As a result, we are able to
investigate the conditions under which mixing between the core and
envelope might occur during the helium flash.  In contrast, standard
calculations skip the helium flash entirely, due to its numerical
complexity, and assume that the only effects of the helium flash are
to remove the degeneracy of the helium core and to increase the carbon
abundance in the core to a few percent by mass (see Sweigart
1994a\markcite{S94a}).  Thus canonical ZAHB models implicitly assume
that the helium flash has no effect on either the envelope mass or
composition.

All of our sequences had an initial helium abundance ($Y$) of 0.23 and
a scaled-solar heavy-element abundance ($Z$) of 0.0015.  During the
first dredge-up along the lower RGB, the envelope helium abundance
increased to 0.244.  As demonstrated by Chieffi, Straniero, \& Salaris
(1991\markcite{C91}) and Salaris, Chieffi, \& Straniero
(1993\markcite{C93}) for the MS and RGB phases, and
additionally by VandenBerg et al.\ (2000\markcite{VSR00}) for the HB
phase, scaled-solar models can closely mimic models with an
enhancement of the $\alpha$ elements, provided the total $Z$ is the same
and the models are not too metal-rich.  For an $\alpha$-element
enhancement of [$\alpha$/Fe]~=~0.3, as is appropriate for metal-poor
globular-cluster stars (Carney 1996\markcite{C96}), this choice for $Z$
corresponds to [Fe/H]~=~$-1.31$, a value within the range of the
metallicity determinations for NGC~2808.

All of our sequences also started with the same main sequence mass $M$
of 0.862~$M_\odot$.  This choice for the mass, together with our
adopted composition, implies an age at the tip of the RGB of 13~Gyr.
VandenBerg (2000\markcite{V00}) has derived a somewhat younger age of
11~Gyr for NGC~2808 from isochrone fitting to the optical CMD after
using the theoretical ZAHB luminosity to set the distance modulus.
Such a modest reduction in the cluster age would, however, have no
effect on the evolutionary behavior to be described below.

Our sequences differ from each other only in the amount of mass loss
along the RGB, which we parameterized with the Reimers
(1975\markcite{R75}, 1977\markcite{R77}) mass-loss formulation:

$\dot{M} = -4 \times 10^{-13} \eta_R L / g R $~~~~($M_\odot$ yr$^{-1}$),

\noindent
where $L$, $g$, and $R$ are the stellar luminosity, gravity, and
radius, respectively, in solar units, and $\eta_R$ is the well-known
Reimers mass-loss parameter.  We considered values of $\eta_R$ from
0.0 (no mass loss) to 1.0.  This range in $\eta_R$ covers the
evolution of stars that ignite helium while on the RGB to those that
ignite helium at high effective temperatures on the WD cooling curve.
For our largest values of $\eta_R$, the models fail to ignite helium,
and instead evolve down the cooling curve as helium WDs.  
Mass loss was terminated in our calculations once the models
evolved off the RGB by 0.1 in log~$\rm T_{eff}$.  At this point
the convective envelope contained $\sim 0.0003~M_\odot$ while the total
envelope mass was $\sim 0.003~M_\odot$.

In the following subsections, we will describe the evolution of these
sequences in more detail, and will demonstrate that models that ignite
helium on the WD cooling curve will undergo substantial flash-induced
mixing between the helium core and the hydrogen envelope, thereby
leading to a natural dichotomy in the subsequent EHB evolution.  The
implications of this flash mixing for the properties of the hottest
EHB stars, the so-called blue-hook stars, will then be explored in the
theoretical plane.  We will also show that models without flash mixing
cannot account for the subluminous EHB stars in NGC~2808 or
$\omega$~Cen.

\subsection{Evolution to the Zero-Age Horizontal Branch} \label{sechb}

Figure~\figeta\ illustrates the various paths taken by our sequences
during their evolution from the MS to the ZAHB.  We define
the ZAHB locus in this figure as the location of our models once the
convective core has stabilized at the beginning of the central
He-burning phase.  The red end of the ZAHB is set by the model
with $\eta_R = 0.0$.

For $\eta_R \lesssim 0.740$, the models remain tightly bound to the RGB
until the helium flash.  A typical example of such evolution is shown
by the track for $\eta_R = 0.620$ (Figure~\figeta $a$). Following
helium ignition, this model rapidly evolved over $\sim 2 \times
10^6$~yr from the tip of the RGB to a ZAHB position on the BHB.  The
track in Figure~\figeta $b$, for $\eta_R = 0.740$, represents the
transition between sequences that ignite helium on the RGB and those
that ignite helium after evolving off the RGB to high effective
temperatures.  In this case, mass loss on the RGB reduced the envelope
mass to $0.03~M_\odot$, and the model arrived on the ZAHB at an
effective temperature of 20,000 K, i.e., just at the boundary between
the BHB and EHB.

Castellani \& Castellani (1993\markcite{CC93}) first showed that stars
that evolve off the RGB due to high mass loss can still undergo the
helium flash.  This result was further explored by D'Cruz et al.\
(1996\markcite{D96}), who demonstrated that a post-RGB helium flash is
possible over a rather large range in $\eta_R$ ($\Delta \eta_R \sim
0.2$).  Two examples of such post-RGB flashes are given in
Figures~\figeta $c$ and \figeta $d$ for $\eta_R = 0.800$ and 0.817,
respectively.  The $\eta_R = 0.800$ sequence ignited helium at a high
luminosity as it was evolving across the theoretical HR diagram, while
the $\eta_R = 0.817$ sequence ignited helium just after it reached the
top of the WD cooling curve.  This latter sequence then evolved to the
hot end of the EHB at an effective temperature of 31,500 K.  As we
shall see, 0.817 is the largest~~value~~of~~$\eta_R$~~for~~which the models
evolved~~to the

%\begin{figure}
%\plotfiddle{fig4.eps}{7.75in}{0}{100}{100}{-250}{45}
%\vskip -0.8in
%\caption{\tiny 
\hskip 0.25in
\parbox{6.5in}{\epsfxsize=6.5in \epsfbox{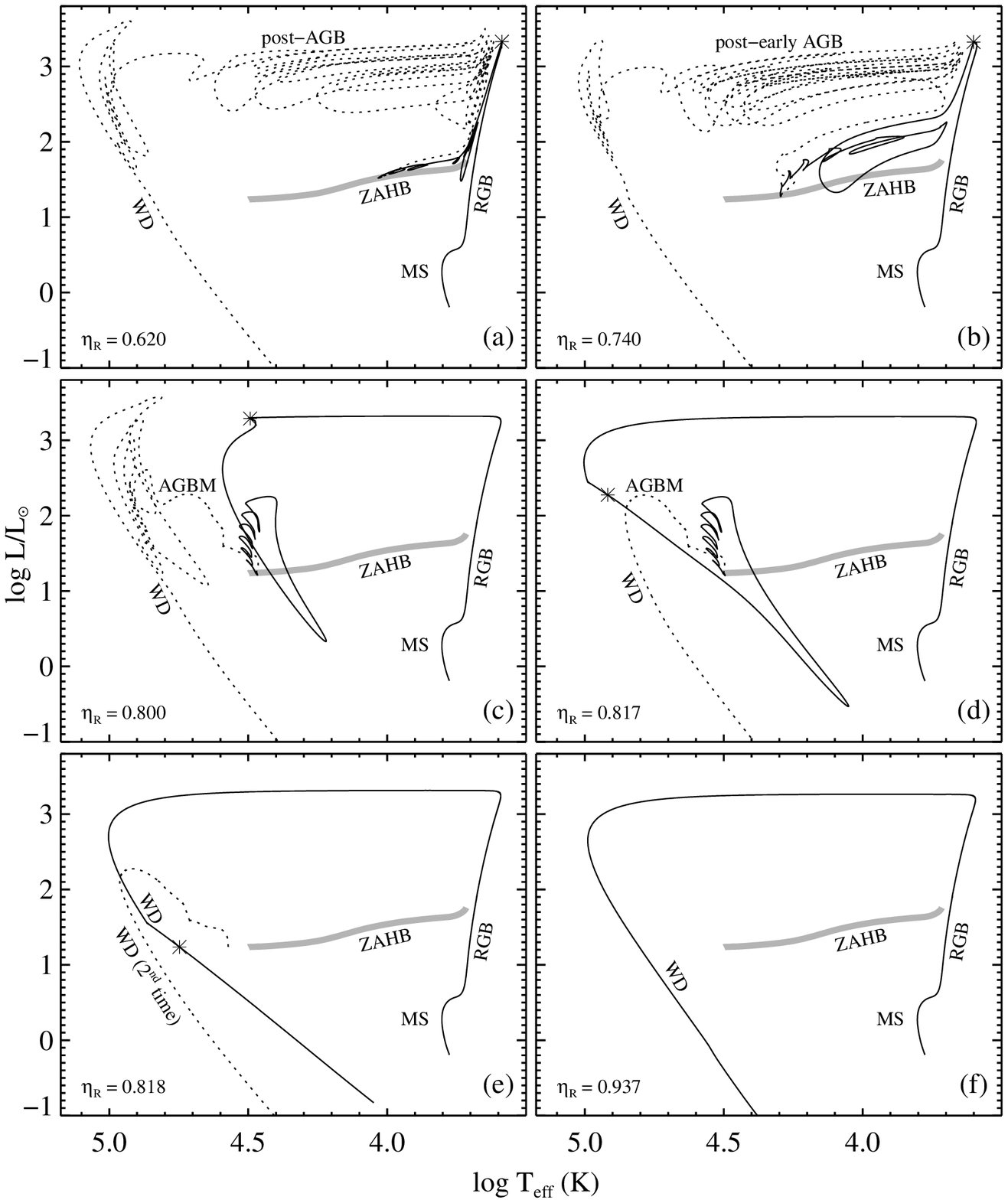}} \\

\vskip 0.1in
\hskip 0.25in
\parbox{6.5in}{\small {\sc Fig.~\figeta--}
Evolutionary tracks for selected values of the Reimers
mass-loss parameter $\eta_R$.  The evolution was followed continuously
through the MS, RGB, and helium-flash phases to the ZAHB (solid lines)
and then through the HB and post-HB phases to the WD cooling curve
(dotted lines).  The post-HB tracks either return to the asymptotic-
giant branch (AGB; panels $a$, $b$) or evolve through an AGB-Manqu$\rm
\acute{e}$ phase (panels $c$, $d$, $e$) before descending the WD
cooling curve. The gyrations in the post-HB tracks are due to helium-
and hydrogen-shell flashes.  As $\eta_R$ increases, the peak of the
main helium-core flash (asterisk) shifts to higher temperatures, and
the subsequent ZAHB location becomes hotter.  The highest value of
$\eta_R$ that avoided flash mixing was $\eta_R = 0.817$ (panel $d$).
The track for $\eta_R = 0.818$ ends when the flash convection
penetrated into the hydrogen envelope (panel $e$).  The subsequent
evolution of this track is represented by a He + C blue-hook
sequence. At $\eta_R > 0.936$ the star will not core-flash at all
(e.g., panel $f$).  }
%\end{figure}

\newpage

%\begin{figure}
%\plotfiddle{fig5.eps}{6.0in}{0}{100}{100}{-250}{50} \vskip -1.0in
%\caption{
\parbox{3.25in}{\epsfxsize=3.25in \epsfbox{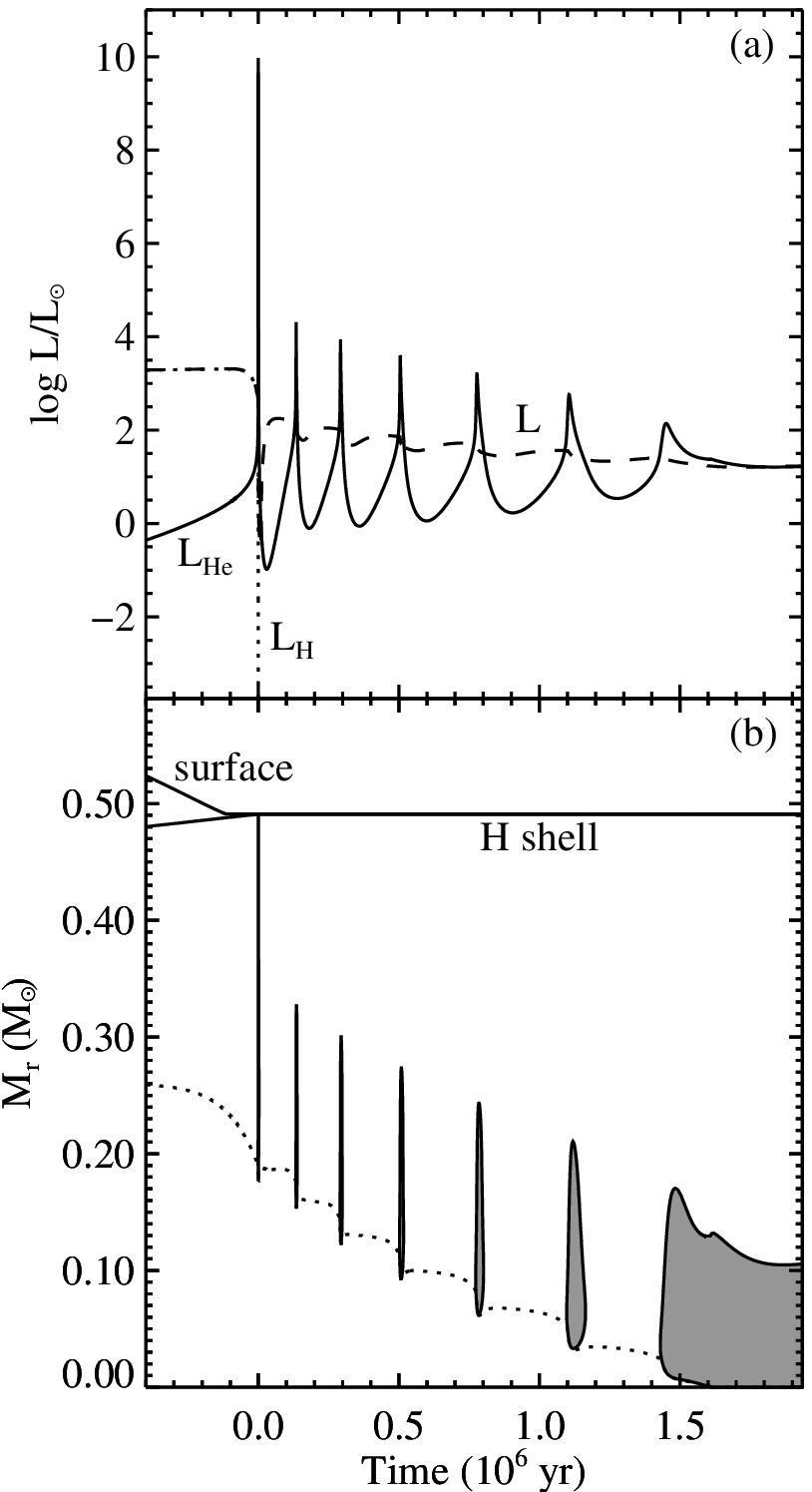}} \\ \\

\parbox{3.25in}{\small {\sc Fig.~\figflash--}
Variation of the interior structure during the evolution through the
helium-core flash to the ZAHB for the sequence with a Reimers
mass-loss parameter $\eta_R = 0.817$.  The zero-point of the timescale
corresponds to the peak of the main flash.  Panel $a$ gives the time
dependence of the helium-burning luminosity (solid curve),
the hydrogen-burning luminosity (dotted curve), and the surface
luminosity (dashed curve).  Panel $b$ gives the location in mass
coordinate $M_r$ of the flash-convection zones (shaded), the center of the
hydrogen shell (labeled), and the stellar surface (labeled).  
The dotted curve in panel $b$ shows the location of the temperature
maximum within the core.  Note that the main flash occurs off-center
at $M_r = 0.18~M_\odot$.  The main flash-convection zone failed to
reach the hydrogen envelope by $\sim 1$ pressure-scale height.  The
evolutionary track for this sequence is plotted in Figure~\figeta $d$.
}
%\end{figure}

\medskip
\medskip

\noindent
ZAHB without flash-induced mixing between the core and
the envelope.

In order to understand better the evolutionary behavior shown in
Figure~\figeta, as well as the conditions required for flash mixing,
we need to look more closely at the structural changes that occur
during the helium flash.  We illustrate these changes for the $\eta_R
= 0.817$ sequence in Figure~\figflash.  This sequence was chosen
because the changes in its interior structure are typical of those for
smaller values of $\eta_R$ and, most importantly, because it lies at
the transition between the sequences with and without flash mixing.
During the main helium-flash at time $t = 0$ in Figure~\figflash $a$,
the He-burning luminosity reached $9.4 \times 10^9 L_\odot$, and the
e-folding time of the thermal instability dropped to only $\sim$1 day.
For 740 yr, the He-burning luminosity exceeded the surface luminosity
($L$) that was present during the previous evolution across the HR
diagram (log~$L / L_\odot \sim 3.3$).  The main flash was then followed
by a series of lower-amplitude secondary flashes, as the He burning
moved inward towards the center.  These secondary flashes are
responsible for the track gyrations that immediately precede the ZAHB
phase in Figure~\figeta.  Eventually the He burning stabilized, and
the star settled onto the ZAHB at $t = 1.94 \times 10^6$~yr.

One might expect the large amount of energy released during the main
flash ($\sim 1.5 \times 10^{49}$~erg) to produce a sudden increase in
the surface luminosity.  However, virtually all of this energy goes
into lifting the degenerate core out of its deep potential well.  In
fact, energy actually flows inward from the inner part of the envelope
into the core just after the main flash peak.  So, even though the star
produces an enormous amount of energy in its deep interior during the
main flash, very little of this energy actually reaches the surface.

The expansion of the core during the main flash cools the hydrogen shell, 
thereby causing the hydrogen-burning luminosity ($L_H$) to drop
abruptly at $t = 0$ in Figure~\figflash $a$.  The star then faces an
energy crises, because it has lost the primary energy source for
supplying its surface luminosity.  The only available energy source
comes from the gravitational contraction of the star's envelope.  The
drop in the surface luminosity following the main flash peak in panels
$a$ to $d$ of Figure~\figeta\ coincides with this envelope
contraction, as the star struggles to fulfill its energy needs.  This
luminosity drop becomes more pronounced with increasing $\eta_R$,
because the envelope mass is then smaller.  The time required for the
surface luminosity to drop from its value at the main flash peak in
Figure~\figeta\ to its subsequent minimum is very short, only
$\sim$8000~yr for the $\eta_R = 0.817$ sequence.

The convection zones produced by the main and secondary flashes during
the $\eta_R = 0.817$ sequence are shown in Figure~\figflash $b$.  Due
to neutrino cooling of the central part of the core during the
preceding RGB phase, the maximum temperature within the core is
located off-center, and consequently the main helium flash occurs in a
shell at the mass coordinate $M_r = 0.18~M_\odot$.  The high outward
flux during the main flash sets up a temporary convection zone that
extends from the flash site out to a point just inside the
hydrogen shell.  This convection zone lasts for only
3,400~yr and is therefore not resolved in Figure~\figflash $b$.  An
expanded view of the outer edge of this convection zone during its
closest approach to the hydrogen shell is presented in
Figure~\figflashzoom.  Note that the flash convection fails to reach
the inner edge of the hydrogen shell by $\sim 10^{-3} M_{\odot}$ or,
equivalently, $\sim 1$ pressure-scale height, and hence flash mixing
does not occur in this case.  Similar results were found for all of
our sequences with $\eta_R \leq 0.817$.

While the main flash removes the degeneracy of the layers outside the flash
site, the layers inside the flash site remain degenerate.  This is simply a
consequence of the fact that very little energy flows into these layers
over the short timescale of the main flash.  The subsequent secondary
flashes then peel off successive layers from this inner degenerate core,
until the temperature maximum and the He burning reach the center and
the star forms a central convective core.  At this point the star has
arrived on the ZAHB.

%\begin{figure}
%\plotfiddle{fig6.eps}{3.25in}{0}{100}{100}{-250}{0}
%\caption{
\parbox{3.25in}{\epsfxsize=3.25in \epsfbox{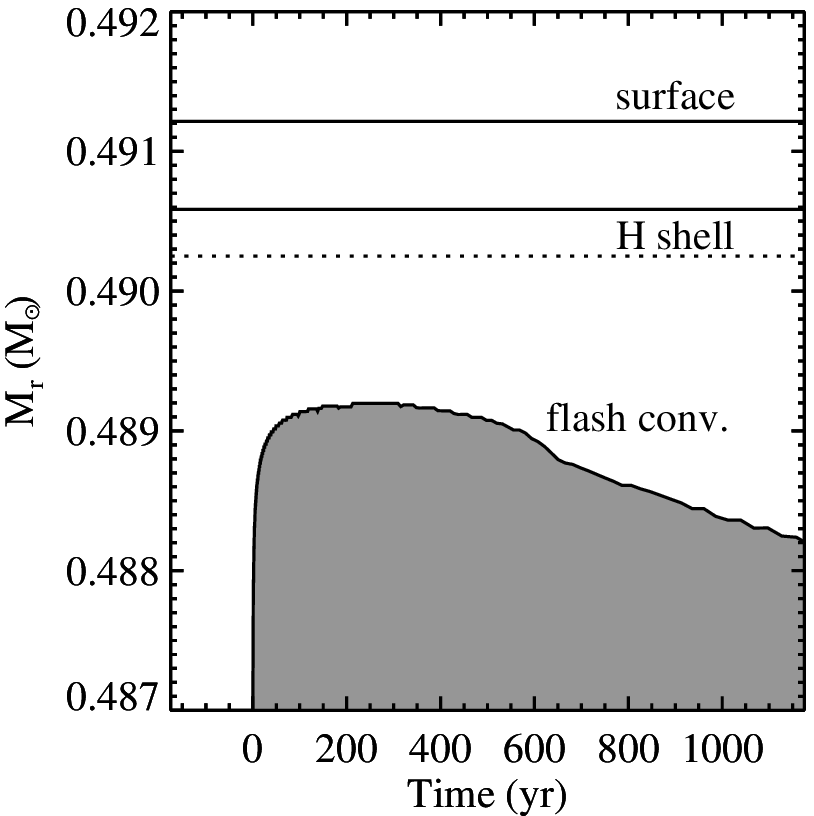}} \\

\parbox{3.25in}{\small {\sc Fig.~\figflashzoom--}
Time dependence of the mass coordinate $M_r$ at the outer edge of the
flash-convection zone, the center of the hydrogen shell, and the
stellar surface during the main helium flash shown in
Figure~\figflash.  The dotted line denotes the inner edge of the
hydrogen shell where $X = 10^{-6}$.
The zero-point of the timescale corresponds to the
peak of the main helium flash.  Shaded areas are convective.
}
%\end{figure}

\medskip

We conclude that flash mixing does not occur if a star ignites helium
either on the RGB or during the evolution to the top of the WD
cooling curve.  In these cases our results confirm the canonical assumption
that the helium flash does not change the envelope mass or composition.

\subsection{Flash Mixing on the White Dwarf Cooling Curve} \label{secflash}

The above evolution changes dramatically when the helium flash occurs
further down the WD cooling curve, as first noted by Sweigart
(1997\markcite{S97}).  Figure~\figeta $e$ shows the evolutionary track
for a sequence where the helium flash did not begin until log~$L / L_\odot
\approx 1.7$.  Following the peak of the main flash at log~$L / L_\odot
\approx 1.2$, the surface luminosity dropped rapidly for 30~yr,
until at log~$L / L_\odot \approx -0.8$ the flash-convection zone
reached the hydrogen shell and began to mix protons from the envelope
into the hot He-burning interior.  At this point the model
calculations were stopped because of the numerical difficulties
discussed below.  The same flash mixing was found for all of our
sequences with $0.818 \leq \eta_R \leq 0.936$.  Sequences with $\eta_R
\geq 0.937$ failed to ignite helium and consequently died as helium
WDs.  An example of this latter evolution is given in Figure~\figeta $f$.

We emphasize the sharpness of the dichotomy between the ``canonical''
evolution for $\eta_R \leq 0.817$ and the ``flash-mixing'' evolution
for $0.818 \leq \eta_R \leq 0.936$.  Even though the difference in
$\eta_R$ between the sequences in Figures~\figeta $d$ and \figeta $e$
is only 0.001, corresponding to a difference in mass loss of only
$10^{-4} M_\odot$, the helium-flash evolution of these two sequences
is strikingly different.  In \S\ref{secbhexp} and \S\ref{secgap} we
will argue that this dichotomy is responsible for both the hot
subluminous stars and the EHB gap in NGC~2808.

The reason for this dichotomy in the helium-flash evolution can be
straightforwardly explained.  What fundamentally distinguishes the
canonical sequences from the flash-mixing sequences is the strength of
the hydrogen shell at the time of the helium flash.  As shown by Iben
(1976\markcite{I76}) in the context of the helium-shell flashes, the
high entropy of a strong hydrogen-burning shell acts as a barrier
preventing the outward penetration of the flash convection into the
hydrogen envelope.  This is the case for all of our sequences with
$\eta_R \leq 0.817$.  However, the size of this entropy barrier
decreases as a star descends the WD cooling curve and its hydrogen shell 
weakens.  Beyond some point, bracketed by the $\eta_R = 0.817$
and 0.818 sequences in Figure~\figeta, flash mixing becomes possible.
Because flash mixing is a consequence of the basic properties of the
stellar models, we conclude that such mixing should be a general
characteristic of any star that ignites helium on the WD cooling
curve.

The mixing found in our helium-{\it core} flash sequences is
remarkably similar to the mixing that occurs during a very late 
helium-{\it shell} flash on the WD cooling curve, according to the
``born-again'' scenario for the origin of the hydrogen-deficient stars
(e.g., Iben et al.\ 1983\markcite{I83}; Iben 1984\markcite{I84},
1995\markcite{I95}; Renzini 1990\markcite{R90}).  In both cases, the
flash-convection zone is only able to penetrate into the hydrogen envelope 
once the hydrogen shell has been partially extinguished.  The
ingestion of the envelope hydrogen by the convection zone during a
very late helium-shell flash leads to s-process nucleosynthesis and
possibly Li production via the Cameron-Fowler mechanism (Herwig \&
Langer 2001\markcite{HL01}; Herwig 2001b\markcite{H01b}).  In
addition, C is dredged up from the deep, He-burning interior.  The
H-deficient, C-rich surface composition resulting from such mixing
supports the ``born-again'' interpretation of the R~CrB stars, the
central stars of planetary nebulae of Wolf-Rayet spectral type (WC),
and the PG~1159 stars (Bl$\rm \ddot{o}$cker 2001\markcite{B01}; Herwig
2001c\markcite{H01c}).  Further support comes from the rapid evolution
and abundance changes observed in the stars FG Sge and Sakurai's
object.  Both of these stars have undergone substantial 
hydrogen-depletion and s-process enrichment, and both seem to be presently
evolving into R~CrB stars (Gonzalez et al.\ 1998\markcite{G98};
Asplund et al.\ 1999\markcite{A99}).  Thus the observational evidence for
flash-induced mixing on the WD cooling curve seems very
strong (Renzini 1990\markcite{R90}; Iben 1995\markcite{I95}).

The theoretical models currently available for the born-again scenario
provide a helpful guide for predicting the effects of flash mixing
during the helium-core flash.  Here we will summarize the main events
that should follow the onset of flash mixing in our evolutionary
sequences.  As protons from the envelope are mixed inward through the
flash-convection zone, they will be carried into regions of higher
temperature and will begin to react with the $\rm ^{12}C$ nuclei
produced by the He burning (Herwig 2001c\markcite{H01c}).  Because
the number of protons in the envelopes of our flash-mixing sequences
is less than the number of $\rm ^{12}C$ nuclei in the flash-convection
zone, these proton-capture reactions will lead primarily to the
production of $\rm ^{13}C$ through the reactions $\rm
^{12}C$(p,$\gamma$)$\rm ^{13}N$($\beta^+$$\nu$)$\rm ^{13}C$ (Sanders
1967\markcite{S67}).  The peak of these reactions will occur around
the layer in the flash-convection zone where the proton-capture
timescale for $\rm ^{12}C$ becomes comparable to the mixing timescale.
Assuming that all of the envelope hydrogen in our flash-mixing models
($5 \times 10^{-4} M_\odot$) is captured by the flash-convection zone,
we find that the energy released by this proton burning will total
$\sim 3\times 10^{48}$ ergs, corresponding to $\sim20$\% of the energy
released by the main helium-flash.  Thus the burning of the envelope
hydrogen will create a substantial, if only temporary, new energy
source for the star.

Two important consequences follow from the development of this new
energy source.  First, the flash-convection zone will split into two
distinct convection zones separated by a thin radiative region, i.e.,
an outer zone powered by proton burning at its base and an inner zone
powered by He burning.  Such splitting of the flash-convection
zone has been found during helium-shell flashes by Sweigart
(1974\markcite{S74}) and Caloi (1990\markcite{C90}).  As a result,
$\rm ^{13}C$ production will be largely confined to the outer
convection zone.  Second, the energy input into the envelope from the
proton burning should expand the star back to giant-branch dimensions,
possibly leading to the formation of a convective envelope and further
dredge-up.  Following the consumption of the envelope hydrogen, these
two convection zones will again coalesce into a single convection
zone.  At this point, $\rm ^{13}C$ from the outer zone will be carried
into the hotter He-burning layers, where s-process nucleosynthesis
via the reaction $\rm ^{13}C$($\alpha$,$n$)$\rm ^{16}O$ can then occur.

The detailed investigation of the flash-mixing phase poses a
formidable numerical challenge, in large part because the
proton-capture reactions within the flash-convection zone occur on a
timescale that is similar to the convective mixing timescale.  Thus
one has to solve for the nucleosynthesis and the time-dependent
convective mixing simultaneously.  In the only available calculation
to follow the evolution of a globular-cluster star through the helium
flash with flash mixing, Sweigart (1997\markcite{S97}) demonstrated
that the flash-convection zone will penetrate deeply into the hydrogen
envelope.  In this case, the subsequent ZAHB model had an envelope
hydrogen abundance $X$ of 0.15, a helium abundance $Y$ of 0.81, and a
carbon abundance of 0.03 by mass.  These calculations did not,
however, include the energy from the proton burning, and therefore
should underestimate the extent of the surface composition changes.
Until recently, all calculations for very late helium-shell flashes in
post-AGB stars likewise omitted the energy from the proton burning
(e.g., Iben et al.\ 1983\markcite{I83}; Iben 1984\markcite{I84}).  The
one exception was the work of Iben \& MacDonald (1995\markcite{IM95}).
To overcome this limitation, Herwig et al.\ (1999\markcite{H99}) and
Herwig (2001a\markcite{H01a}) have developed a new diffusion algorithm
for coupling the nucleosynthesis to the convective mixing, and have
successfully applied their algorithm to the problem of mixing during a
very late helium-shell flash.  Their results, as well as those of Iben
\& MacDonald (1995\markcite{IM95}), show that flash-induced mixing will
strongly deplete the envelope hydrogen abundance.  It would be
especially interesting to apply this technique to the case of a
helium-core flash, although the numerics would probably be more
demanding, due to the higher burning rates and shorter evolutionary
timescale.

In view of the above numerical difficulties, we terminated our
sequences as soon as they encountered flash mixing.  Nevertheless, we
can still predict the surface composition that these sequences should
have when they arrive on the ZAHB.  The calculations discussed above
clearly indicate that flash mixing will consume most, if not all, of
the envelope hydrogen.  Thus we expect the surface composition of the
flash-mixing models to be strongly depleted in hydrogen and enriched
in helium.  The surface carbon abundance should be close to its value
in the flash-convection zone, namely, 0.04 by mass.  In addition, an
enhancement of the s-process elements, as seen in FG~Sge and Sakurai's
object, would also be expected.

In the following subsection, we will explore the impact of these composition
changes on the HB evolution and, more specifically, will show that these
changes will lead to a dichotomy in the ZAHB effective temperature.  Both
the differences in the surface composition and the ZAHB effective
temperature will prove important for understanding the properties of the
blue-hook stars.  We will also contrast our ``flash-mixing'' models with
models that ignore the effects of flash mixing.

\subsection{Blue-Hook Models with Hydrogen-Depleted Envelopes} \label{secbh}

As explained above, there is a clear distinction between our sequences
with $\eta_R \leq 0.817$ and those with $0.818 \leq \eta_R \leq
0.936$.  For $\eta_R \leq 0.817$, the models undergo the helium flash
prior to descending the WD cooling curve and, as a result, evolve to
the canonical ZAHB without any flash-induced mixing between the
helium core and hydrogen envelope.  In contrast, the models for $0.818
\leq \eta_R \leq 0.936$ do not ignite helium until they are on the WD
cooling curve.  Under such circumstances the flash-convection zone is
able to penetrate deeply into the envelope, thereby greatly modifying
the envelope mass and composition.  As will be seen below, the models
with flash mixing form a ``blue-hook'' feature near the hot end of the
EHB as they begin their HB evolution.  Such a feature is evident in
the CMD of $\omega$~Cen reported by D'Cruz et al.\
(2000\markcite{D00}).  In \S\ref{secbhexp} we will identify both these
blue-hook stars in $\omega$~Cen and the hot subluminous stars in
NGC~2808 with our EHB models that undergo flash-mixing.  For these
reasons we will refer to our EHB models with flash mixing as
``blue-hook'' models in the following discussion.  This will also
serve to differentiate them from the canonical EHB models without
flash mixing.

Some properties of our sequences with $\eta_R \geq 0.60$,
corresponding to ZAHB $\rm T_{eff} \gtrsim 10,000~K$, are presented in
Figures \figetatemp\ and \figetamass.  As indicated in
Figure~\figetatemp, the canonical EHB and blue-hook regions are
populated over a rather large range in $\eta_R$, from 0.740 to 0.936.
As mentioned earlier, this result agrees well with the results of
D'Cruz et al.\ (1996\markcite{D96}), who found that EHB models could
be produced over a similarly large range ($\Delta \eta_R \sim 0.2$)
for both metal-poor and metal-rich compositions.  Note that the EHB
models of D'Cruz et al.\ (1996\markcite{D96}) include both the
canonical EHB and blue-hook models in Figure~\figetatemp.  The lower
panel of Figure~\figetatemp\ shows, however, that only models within
the narrow range from $0.780 \leq \eta_R \leq 0.817$ ignite helium
while evolving from the tip of the RGB to high effective temperatures.
Most of the hot ZAHB models in the top panel of Figure~\figetatemp\
ignite helium either as they are just peeling off the RGB ($0.740 \leq
\eta_R \leq 0.780$) or as they are descending the WD cooling curve
($0.818 \leq \eta_R \leq 0.936$).  Indeed, the blue-hook models span
$\sim$60\% of the range in $\eta_R$ producing ZAHB models hotter than
20,000~K.  Given a uniform distribution in $\eta_R$, we would
therefore expect somewhat more than half of the ``hot He-flashers''
discussed by D'Cruz et al.\ (1996\markcite{D96}) to undergo flash
mixing.  This is also evident in Figure 1 of D'Cruz et al.\
(1996\markcite{D96}), where many of the hot He-flashers lie on the WD
cooling curve at helium ignition.

The variation in the total mass ($M$), core mass ($M_c$), and envelope
mass ($M_{env}$) with $\eta_R$ is shown in Figure~\figetamass\ for our
BHB, EHB, and blue-hook models.  The total mass decreases linearly with
$\eta_R$ until the models begin to peel away from the RGB prior to
helium ignition.  The change in slope at $\eta_R \approx 0.79$ is a
consequence of turning off the Reimers mass-loss in the calculations
after the models leave the RGB.  As expected, the core mass is
virtually constant at $M_c = 0.491~M_\odot$ for $\eta_R \leq 0.817$,
but then decreases by $\sim0.01~M_\odot$ for the blue-hook models.  In
the next subsection, we will demonstrate that such a modest decrease
in $M_c$ cannot (by itself) explain the subluminous stars in NGC~2808
and $\omega$~Cen.  While one might expect the envelope mass to
decrease monotonically with $\eta_R$, this is not, in fact, the case.
The bottom panel of Figure~\figetamass\ shows that $M_{env}$ reaches a
lower limit of $\sim6 \times 10^{-4} M_\odot$ for all of the blue-hook
models.  The same behavior is also found in Table 1 of D'Cruz et al.\
(1996\markcite{D96}), where the minimum $M_{env}$ decreases from
$\sim2 \times 10^{-3} M_\odot$ at [Fe/H]~$= -2.26$ to $\sim 8 \times
10^{-4} M_\odot$ at [Fe/H]~=~0.37.  This result implies that canonical
ZAHB models cannot be extended to arbitrarily small $M_{env}$ in order
to match the hottest observed HB stars, as has sometimes been assumed
(see, e.g., Sosin et al.\ 1997\markcite{SDD97}).  We will show below
that the composition changes associated with flash-mixing of the
envelope naturally create hotter stars, thus alleviating the need for
very small envelope masses.

As discussed previously, our blue-hook sequences were not evolved
completely through the helium-flash phase, due to the numerical
difficulties associated with the flash mixing, and consequently we had
to adopt a different procedure for computing the blue-hook ZAHB
models.  We proceeded in two steps.  Starting with the canonical ZAHB
model for $\eta_R = 0.817$, we first adjusted the values of $M_c$ and
$M_{env}$ until they matched the values for the blue-hook models
plotted in Figure~\figetamass.  A similar procedure is often used to
compute canonical ZAHB models with different envelope masses.  Next we
changed the envelope composition of the models.  Because detailed
calculations including the energetics are not yet available for the
flash-mixing phase, we decided to construct three sets of blue-hook
models in order to explore the effects of different envelope
compositions.  The first set (hereafter He+C models) had an enhanced
helium abundance of 0.96 and a carbon abundance of 0.04 by mass, with
the remaining heavy elements having their initial cluster abundances.
As noted above, such a composition reflects the composition of the
flash-convection zone and is therefore the most~~likely~~outcome of
flash mixing.  The second~~set 

\vskip 0.1in

%\begin{figure}
%\plotfiddle{fig7.eps}{6.0in}{0}{100}{100}{-250}{0}
%\caption{ 
\parbox{3.25in}{\epsfxsize=3.25in \epsfbox{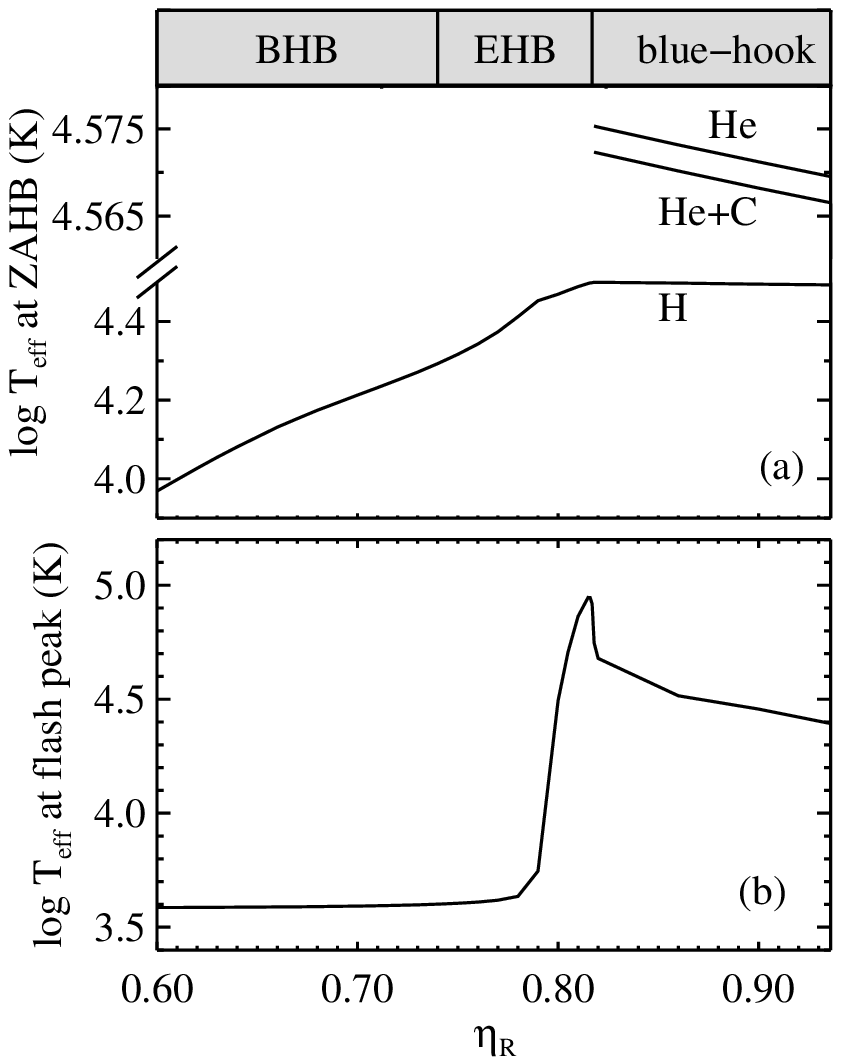}} \\

\parbox{3.25in}{\small {\sc Fig.~\figetatemp--}
Effective temperature at the ZAHB (panel $a$) and at the
peak of the main helium flash (panel $b$) as a function of the Reimers
mass-loss parameter $\eta_R$.  At the top of the figure we indicate
the ranges in $\eta_R$ giving rise to the BHB, canonical EHB, and
blue-hook models.  The log~$\rm T_{eff}$ scale changes at log~$\rm
T_{eff} = 4.50$ in panel $a$ in order to show the differences in the
predicted temperatures of the H, He, and He+C blue-hook models more
clearly.}
%\end{figure}

\medskip

%\begin{figure}
%\plotfiddle{fig8.eps}{4.32in}{0}{100}{100}{-250}{-20}
%\caption{
\parbox{3.25in}{\epsfxsize=3.25in \epsfbox{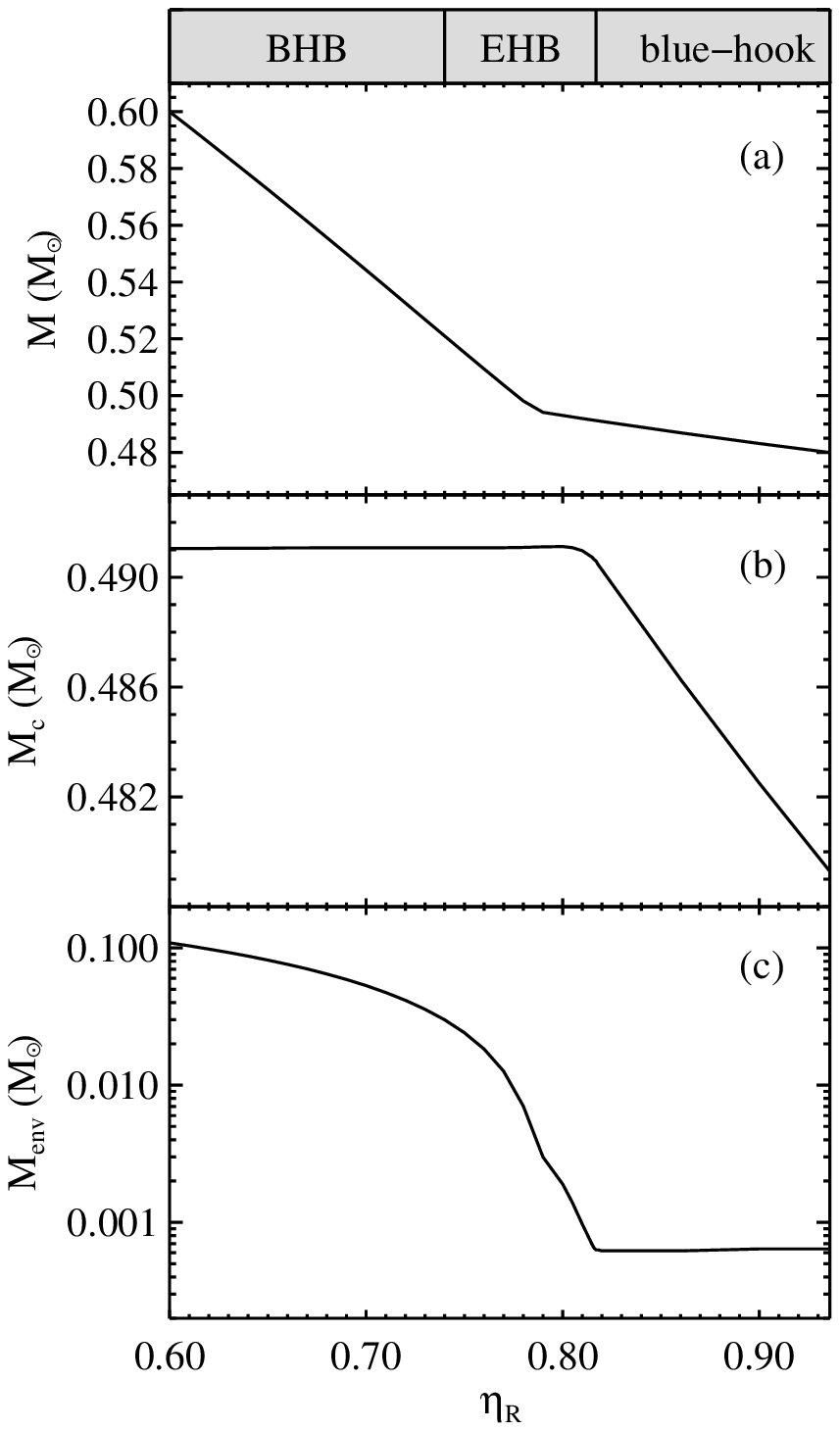}} \\

\parbox{3.25in}{\small {\sc Fig.~\figetamass--}
Total mass $M$ (panel $a$), core mass $M_c$ (panel $b$), and envelope
mass $M_{env}$ (panel $c$) at the helium flash as a function of the
Reimers mass-loss parameter $\eta_R$.  At the top of the figure we
indicate the ranges in $\eta_R$ giving rise to the BHB, canonical EHB, and
blue-hook models.  The core mass is constant for $\eta_R \lesssim 0.80$
and then decreases by $\sim 0.01~M_\odot$ for the blue-hook models.
Note that all of the models that ignite helium on the WD cooling curve
have nearly the same envelope mass of $6 \times 10^{-4} M_\odot$.}
%\end{figure}

\vskip 0.1in

\noindent
(hereafter He models) had a pure helium
envelope, except for the initial heavy-element abundance of 0.0015.
With this set we could determine the effects of enhanced helium by
itself.  Finally the third set (hereafter H models) had the same
hydrogen-rich envelope composition as the canonical EHB models.  We
computed this set to determine where blue-hook models without flash
mixing would lie in a CMD.  Discussion of these H blue-hook models
will be postponed to the next subsection.  All of these calculations
were carried out for four values of $\eta_R$, namely, $\eta_R$ =
0.818, 0.860, 0.900 and 0.936.  We then evolved all of these ZAHB
models through the HB and post-HB phases.

The HB evolutionary tracks for each set of blue-hook models, together
with some of our hottest canonical EHB tracks, are plotted in
Figure~\figbhevol.  The rate of evolution along these tracks is
indicated by the dots, which are separated by a time interval of
$10^7$~yr.  Overall these tracks have the expected morphology, namely,
they evolve upward in luminosity at an approximately constant $\rm
T_{eff}$.

%\begin{figure}
%\plotfiddle{fig9.eps}{7.25in}{0}{100}{100}{-250}{40}
%\vskip -0.9in
%\caption{
\parbox{6.5in}{\epsfxsize=6.5in \epsfbox{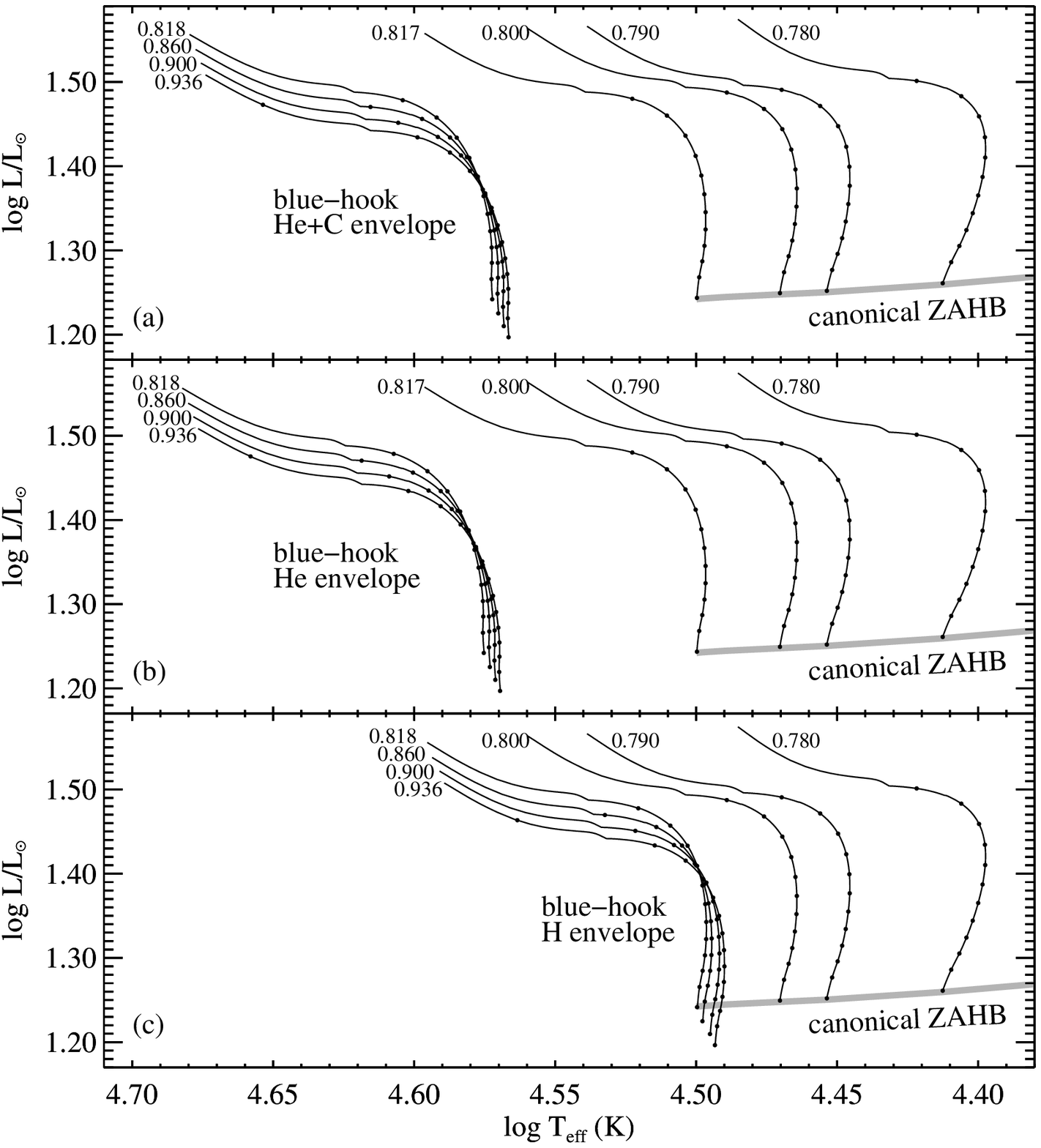}} \\ 

\hskip 0.25in
\parbox{6.5in}{\small {\sc Fig.~\figbhevol--}
HB evolutionary tracks for blue-hook sequences with He+C envelopes
(panel $a$), He envelopes (panel $b$) and H envelopes (panel $c$).
Each track is labeled by its value of 
$\eta_R$.  For comparison, we plot the canonical HB tracks for $\eta_R
= 0.780$, 0.790, 0.800, and 0.817, as well as the canonical ZAHB (grey
curves).  The dots along each track are separated by a time interval
of $10^7$ yr.  Note the dichotomy between the predicted effective
temperatures for the He+C and He blue-hook tracks and the canonical
tracks.  In contrast, the H blue-hook tracks lie at the hot end of the
canonical EHB.  }
%\end{figure}

\medskip

Several features of the tracks in Figure~\figbhevol\ deserve comment.
First, we note that there is a well-defined high temperature limit to
the canonical EHB at $\rm T_{eff} \approx 31,500~K$.  Within the
canonical framework, it is not possible to produce hotter EHB stars,
because the envelope mass cannot be reduced below $6\times 10^{-4} M_\odot$ 
for $Z = 0.0015$, regardless of the extent of mass loss along the RGB.  In
contrast, the blue-hook tracks with He+C or He envelopes form a
distinct group located at a significantly higher effective temperature
than the hot end of the canonical EHB.  The~~properties~~of~~the~~
blue-hook~~tracks~~within~~each~~group \\

\vskip 7.95in

\noindent
depend only slightly on the value
of $\eta_R$.  They are also rather insensitive to the envelope carbon
abundance, although some differences will be apparent when these
tracks are transformed to the observational plane (see \S\ref{secbhexp}).

Figures \figbhevol $a$ and \figbhevol $b$ again illustrate the
abruptness with which the tracks shift blueward with the onset of
flash mixing (see also Figure \figetatemp $a$).  Over an interval of
only 0.001 in $\eta_R$, from 0.817 to 0.818, the predicted HB
temperature jumps by $\sim 6000$~K, from $\approx 31,500$~K to
$\approx 37,200$~K.  For comparison, the same change in $\eta_R$ near
the hot end of the canonical EHB would produce an increase in $\rm
T_{eff}$ of only $\sim 100$~K.  These results are consistent with
the previous calculations of Sweigart (1997\markcite{S97}), who also
found that his flash-mixing model did not lie at the hot end of the
canonical EHB, but rather at a much higher effective temperature,
closer to the domain of the high gravity He-sdO stars (Lemke et al.\
1997\markcite{L97}).  It is also interesting to note that the only
known He-sdB star in a globular cluster (F2-2 in M15) has an effective
temperature of 36,000~K (Moehler, Heber, \& Durrell
1997\markcite{M97}), close to that predicted for our blue-hook models
with flash mixing.

These features of the blue-hook tracks with He+C or He envelopes have
some important implications.  First, it is clear that flash mixing
introduces a natural dichotomy in the properties of hot HB stars,
which manifests itself as a gap centered at $\rm T_{eff} \approx
34,400$~K in the theoretical temperature distribution.  In
\S\ref{secgap} we will identify this flash-mixing gap with the high
temperature gap observed within the EHB of NGC~2808.  At first glance,
there would appear to be a mismatch between this theoretical gap
temperature and the temperature of $\sim 25,000$~K derived by fitting
canonical ZAHB models to the optical CMD of NGC~2808 (Sosin et al.\
1997\markcite{SDD97}; Bedin et al.\ 2000\markcite{BPZ00}).  However,
this apparent mismatch can be largely attributed to HB evolution.  Due
to the temperature insensitivity of $B - V$ along the HB blue tail,
the tracks in Figure~\figbhevol\ will evolve vertically towards
brighter $V$ magnitudes in the ($V$, $B - V$) CMD.  Thus the blue-hook
tracks will appear to fill in the range in $V$ magnitudes corresponding
to the temperature range of the flash-mixing gap.  At the same time,
the hottest EHB models will also evolve towards brighter $V$ magnitudes,
thereby depleting the hot end of the canonical EHB.  The net effect is
to shift the apparent location of the EHB gap towards cooler
temperatures than the actual gap in the stellar parameters.  This
point is discussed more fully in \S\ref{secgap}.

Due to their smaller core masses, the blue-hook ZAHB models in Figures
\figbhevol $a$ and \figbhevol $b$ lie somewhat below the extension of
the canonical ZAHB.  However, these ZAHB models are only fainter by
$\sim 0.1$~mag, whereas the hot subluminous HB stars in NGC~2808 and
$\omega$~Cen lie as much as $\sim 0.7$~mag below the ZAHB in the UV.
In \S\ref{secbhspec} and \S\ref{secbhexp} we will show that the
hydrogen-depleted surface compositions of the flash-mixing models will
suppress the far-UV flux in the stellar spectra and thereby
potentially explain the fainter luminosities of these sub-ZAHB stars.
The higher effective temperatures of the He+C and He blue-hook models
will also imply larger bolometric corrections, and hence fainter $V$
magnitudes, for these models in an optical CMD.  Thus, the blue-hook
and canonical EHB models make different predictions concerning the
faint end of the blue HB tail.  These predictions will be tested
against the observed faint end of the HB in NGC~2808 (Walker
1999\markcite{W99}; Bedin et al.\ 2000\markcite{BPZ00}), once our
models have been transformed to the observational plane.

\subsection{Blue-Hook Models With Hydrogen-Rich Envelopes} \label{secbhh}

We now consider the blue-hook models with hydrogen-rich envelopes
plotted in Figure~\figbhevol $c$.  These models assume that flash mixing 
does not change either the envelope mass or composition, and
therefore they represent, in effect, the continuation of the canonical
EHB to higher mass-loss rates.  We computed these H blue-hook models
in order to study how hot HB models evolve when flash mixing is
ignored, and to provide a set of hydrogen-rich models for comparison
with the He+C and He blue-hook models.

The properties of the H blue-hook models differ significantly in
several respects from those of the He+C and He blue-hook models.  We
first note that all of the H blue-hook models in Figure~\figbhevol $c$
lie near the hot end of the canonical EHB, forming a hook-like feature
that extends to slightly cooler temperatures and fainter luminosities.
There is, however, no dichotomy in the predicted effective
temperatures between the H blue-hook models and the canonical EHB
models, and hence no obvious way to produce the EHB gap observed in
NGC~2808.  Moreover, there is no increase in the maximum EHB
temperature beyond the hot end of the canonical EHB, and thus no
increase in the length of the blue HB tail in the optical CMD.  All of
this is a consequence of the very similar envelope masses of the H
blue-hook models and the hottest canonical EHB models.  The fainter
luminosities of the H blue-hook models are due entirely to the smaller
core masses of these models.  However, the maximum luminosity offset
from the canonical ZAHB is only $\sim 0.1$~mag, much less than
observed among the hot sub-ZAHB stars in NGC~2808 and $\omega$~Cen.
This discrepancy cannot be explained by differences in the spectral
energy distribution, because both the H blue-hook and canonical EHB
have the same hydrogen-rich envelope composition.

The H blue-hook models in Figure~\figbhevol $c$ can be directly
compared to the ``hot He-flashers'' of D'Cruz et al.\
(1996\markcite{D96}, 2000\markcite{D00}), which likewise do not
include the effects of flash mixing.  As can be seen in Figure 2 of
D'Cruz et al.\ (1996\markcite{D96}), the hot He-flashers all lie at
the hot end of the canonical EHB, with no indication of a gap in the
temperature distribution between these models and the canonical EHB
models.  Again we find that the luminosities of the hot He-flashers
are, at most, only $\sim 0.1$~mag fainter than the canonical ZAHB, due
to the fact that the core masses of these models are only
$\sim 0.015~M_\odot$ smaller than the canonical value.  Overall the H
blue-hook models and hot He-flashers agree very well.

The only way in which the H blue-hook models and the hot He-flashers
can account for the faint sub-ZAHB stars in NGC~2808 and $\omega$~Cen
is by reducing the mass of their helium cores.  To examine this point
more closely, we computed a series of HB sequences starting with a
representative EHB model with $\eta_R = 0.780$.  In these sequences,
we reduced the core mass in increments of $0.01~M_\odot$ to a value
$0.08~M_\odot$ smaller than the canonical value, while keeping the
envelope mass constant at $0.007~M_\odot$.  The results, shown in
Figure~\figredcore, demonstrate that one would have to reduce $M_c$ by
$\sim 0.06~M_\odot$ in order to lower the ZAHB luminosity by 0.7~mag,
as observed.  The same conclusion can also be obtained from the HB
sequences of Sweigart \& Gross (1976\markcite{SG76}). Such a large
reduction in $M_c$ is completely inconsistent with current
evolutionary calculations (Sweigart 1994b\markcite{S94b}).  Even if
such a reduction were possible, one would still face the quandary of
understanding why only some stars had such small core masses.

We conclude that hot HB models with hydrogen-rich envelopes such as
the present H blue-hook models or the hot He-flashers of D'Cruz et
al.\ (1996\markcite{D96}) cannot explain the faint UV luminosities of
the hot sub-ZAHB stars in NGC~2808 or $\omega$~Cen.  In
\S\ref{secbhexp} we will show that such faint luminosities might be
explained if the stellar envelope is hydrogen-deficient.  The sub-ZAHB
stars would then be fainter because of a difference in their spectral
energy distribution, which suppresses the far-UV flux, as well as an
increase in the bolometric correction.

\section{MODEL ATMOSPHERES AND SYNTHETIC SPECTRA} \label{secspec}

To translate the stellar evolutionary models in \S\ref{secevol} to the
observational plane, we used several different sources of model
atmospheres and synthetic spectra, depending upon the type of star.
It is extremely time-consuming and resource-intensive to compute new
models, so we used existing grids of atmospheres and spectra where the
accuracy of these older grids sufficed.  New models were computed when
an existing model could not accurately describe a given type of star.
We discuss the models below.

\subsection{Horizontal Branch Stars} \label{sechbspec}

For the canonical HB sequences, we used synthetic spectra at the
cluster metallicity.  These synthetic spectra were all calculated
under the assumption of local thermodynamic equilibrium (LTE), which
means that the distribution of atoms among their excitation and
ionization states was calculated from the local values of two
thermodynamic variables: temperature and electron density.  At
log~$g\le 5.0$ (e.g., $\rm T_{eff} < $~20,000~K on the ZAHB), we used
the Kurucz (1993\markcite{K93}) grid of synthetic spectra,
interpolating in $\rm T_{eff}$ and metallicity from the grid points
that most closely matched each HB model.  At HB surface gravities
exceeding those available in the Kurucz (1993\markcite{K93}) grid, we
used the ATLAS9 model atmosphere program (Kurucz 1993\markcite{K93})
to generate new LTE model atmospheres at the $\rm T_{eff}$ and surface
gravity for each HB model, with metallicities bracketing the cluster
metallicity (we can only run ATLAS9 at the discrete metallicity steps
available in the published Kurucz grid).  Note that most researchers
simply use the Kurucz spectra at log~$g=5$ to derive the colors of HB
stars that exceed this gravity, and this practice is acceptable,
because the change in surface gravity has only a small effect on the
broad characteristics of the spectrum (the surface gravity on the ZAHB
does not exceed log~$g=6$).  We generated these new atmospheres at
higher gravities to ensure that a gravity mismatch on the EHB would
not be the cause of any discrepancies between the data and stellar
theory.  

\medskip

%\begin{figure}
%\plotfiddle{fig10.eps}{3.5in}{0}{100}{100}{-250}{0}
%\caption{
\parbox{3.25in}{\epsfxsize=3.25in \epsfbox{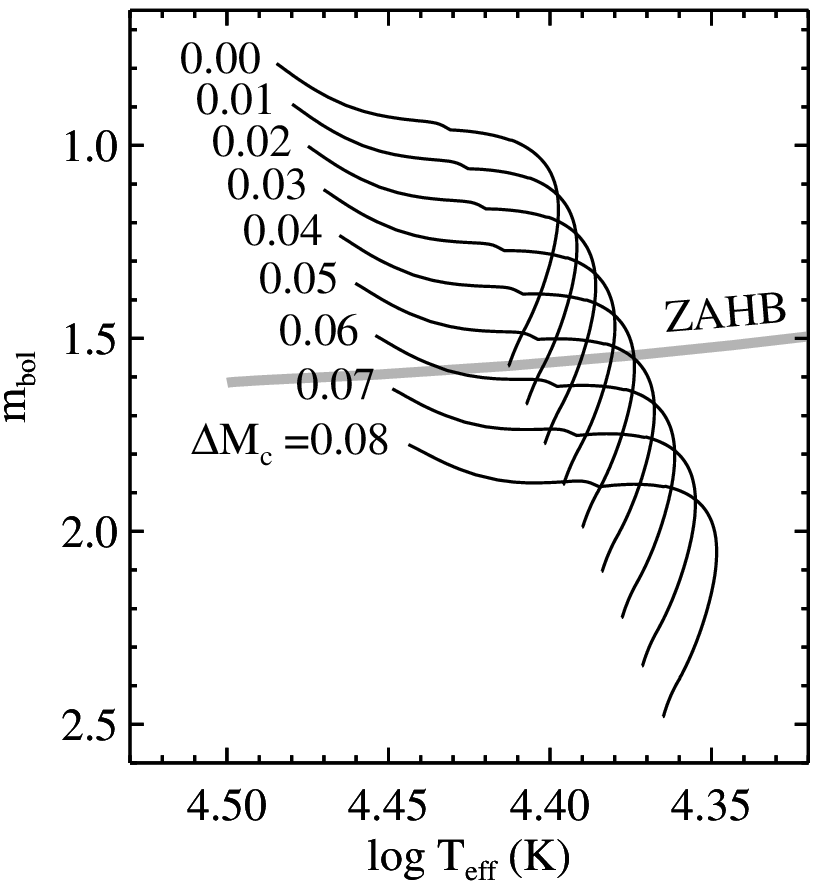}} \\

\parbox{3.25in}{\small {\sc Fig.~\figredcore--}
Dependence of the bolometric magnitude along the HB evolutionary track
of a representative EHB model on the mass of the helium core.  The
end of each track is labeled by the amount $\Delta M_c/M_\odot$ by
which the core mass has been reduced below its canonical value.  Note
that an implausibly large decrease of $\Delta M_c \sim 0.06~M_\odot$
would be required to explain the luminosities of the hot sub-ZAHB
stars in NGC~2808 and $\omega$~Cen.}
%\end{figure}

\noindent
We then used the SYNSPEC code (Hubeny et al.\
1994\markcite{HLJ94}) to create synthetic spectra from these model
atmospheres, and the spectra were interpolated to give synthetic
spectra at the cluster metallicity.  Note that the agreement between
the SYNSPEC spectra and the Kurucz spectra, where they meet at
log~$g$~=~5 on the HB, is excellent.
\subsection{Blue-Hook Stars} \label{secbhspec}

Synthetic spectra with scaled solar abundances would not accurately
represent the unusual stellar atmospheres present in blue-hook stars,
because these stars should have greatly enhanced helium and carbon
(see \S\ref{secbh}).  The Kurucz (1993\markcite{K93}) grid assumes
scaled-solar abundances, so we generated new model atmospheres with
independently scaled abundances for the individual elements, using the
TLUSTY model atmosphere code (Hubeny \& Lanz 1995\markcite{HL95}).  We
calculated atmospheres for each set of blue-hook models, using the
three envelope compositions explained in \S\ref{secbh} and displayed
in Figure~\figbhevol.  The first composition set assumed enhanced
helium (0.96 by mass) and enhanced carbon (0.04 by mass), with all
other elements at the cluster abundance.  The second set assumed
enhanced helium (nearly 1.0 by mass) with heavier elements at the
cluster abundance.  The third set assumed scaled-solar abundances for
all elements, with [Fe/H]~=$-1.36$.  We also computed two variations
of the first set.  One of these variations replaced the enhanced
carbon with enhanced nitrogen, in case the carbon in the envelope was
burned to nitrogen during the flash mixing and nucleosynthesis
(although we note that there should not be enough protons in the
envelope to do so).  The other variation enhanced only the iron abundance,
from the nominal cluster abundance, to a value of 1.25\% by mass (i.e.,
10 times the solar Fe mass fraction), to simulate the effects of
radiative levitation on iron.

Hydrogen, helium, carbon, nitrogen, and iron were allowed to depart
from LTE. About 550 individual levels of H, He, C, and N, and about
43,000 Fe levels, of the following ions, were included in the non-LTE
TLUSTY model atmospheres: \ion{H}{1}, \ion{He}{1}, \ion{He}{2},
\ion{C}{2}, \ion{C}{3}, \ion{C}{4}, \ion{N}{2}, \ion{N}{3},
\ion{N}{4}, \ion{N}{5}, \ion{Fe}{3}, \ion{Fe}{4}, \ion{Fe}{5},
and \ion{Fe}{6}. These levels were grouped into about 400 superlevels. At
the cluster abundance, the iron line-blanketing effect remains very
small, and the atmosphere structure is significantly changed only when
the iron abundance is strongly enhanced.  Thus, we explicitly included
the non-LTE treatment of Fe in our model atmospheres only when Fe was
enhanced.  When the atmosphere is carbon-rich, the details of the
carbon opacity also become important, in particular strong and broad
autoionization features appear in the far-UV spectrum.  This opacity
comes from the detailed cross-sections of the Opacity Project (Cunto
et al.\ 1993\markcite{C93}), along with our use of the TLUSTY Opacity
Sampling mode with small frequency steps.

Once we calculated the model atmospheres, we used the SYNSPEC code
(Hubeny et al.\ 1994\markcite{HLJ94}) to generate non-LTE synthetic
spectra for each atmosphere.  Species not included explicitly in the
TLUSTY models were assumed to be in LTE.  These spectra covered the
range of 1,000--10,000~\AA\ at 1~\AA\ resolution, to allow computation
of observed colors in various HST and ground-based bandpasses.

Our calculations show that blue-hook stars with flash-mixed
atmospheres have significantly different spectra than stars at the
cluster metallicity.  We show the effects of these abundance
enhancements in Figure~\figatmos.  Note that for simplicity, this
figure shows the emergent flux from the model
atmospheres, and not the detailed synthetic spectra.  In a normal
stellar atmosphere composed mostly of hydrogen, the hydrogen opacity
shortward 

%\begin{figure}
%\plotfiddle{fig11.eps}{6.0in}{0}{100}{100}{-250}{45}
%\vskip -0.9in
%\caption{
\hskip 0.25in
\parbox{6.5in}{\epsfxsize=6.5in \epsfbox{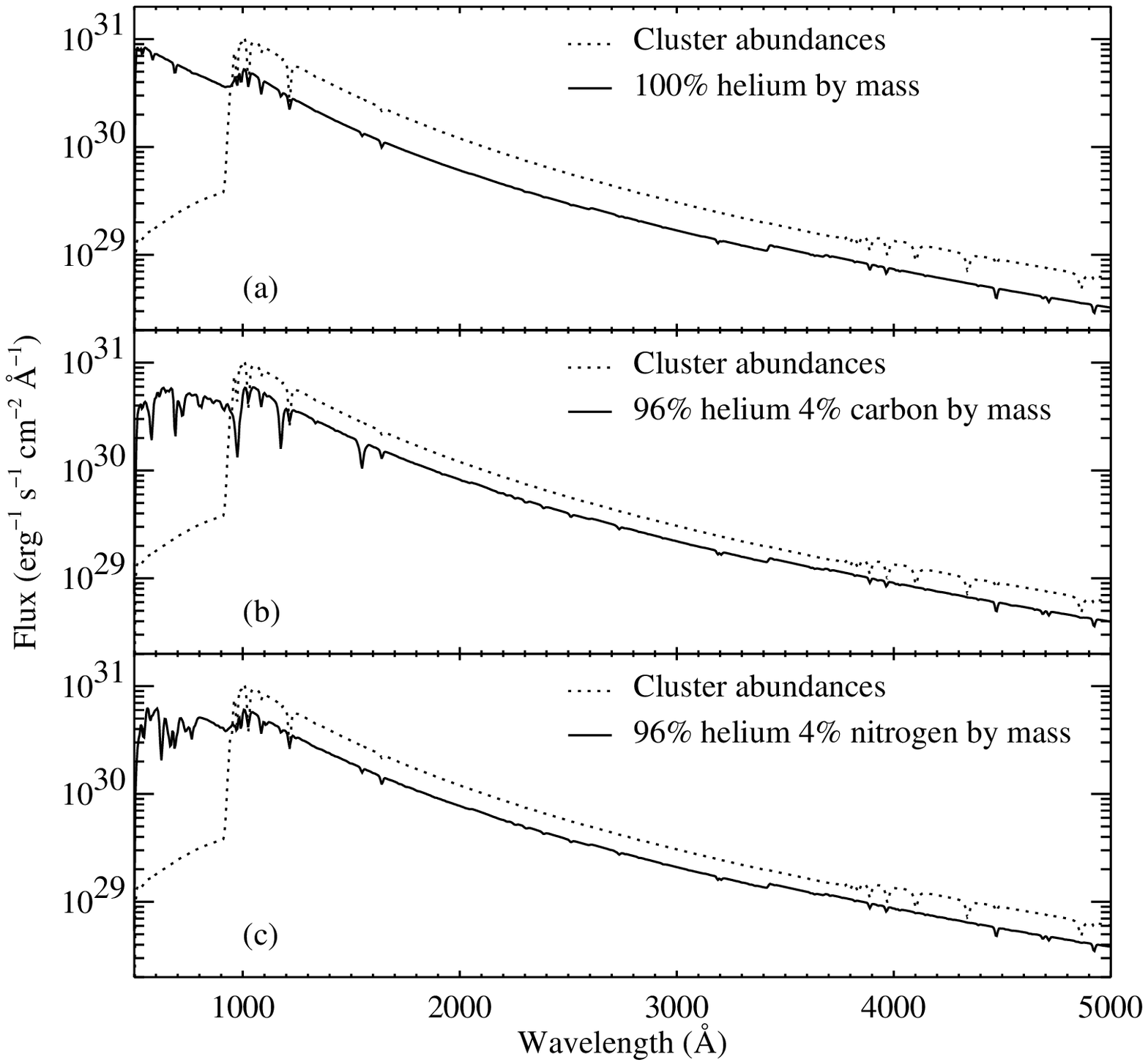}} \\ 

\hskip 0.25in
\parbox{6.5in}{\small {\sc Fig.~\figatmos--}
Spectral energy distributions for a blue-hook star (solid
curves) near $\rm T_{eff} = 37,000$~K, for different assumed envelope
compositions (see \S\ref{secbh}).  In each panel, we show the emergent
flux (not the detailed synthetic spectrum) on a log scale.  The dashed
line in each panel shows the flux from a canonical EHB atmosphere
($\rm T_{eff} = 31,000~K$) that assumes the nominal cluster
abundances.  Panel $a$ shows the flux from an atmosphere that is
nearly 100\% helium by mass; the reduction in hydrogen opacity
increases the flux below 912~\AA, at the expense of flux at longer
wavelengths.  In panel $b$, increasing the carbon abundance restores
some of the sub-912~\AA\ opacity, and so the flux emerges at longer
wavelengths. In panel $c$, we show that nitrogen could also provide
this opacity, and thus the spectrum is similar to that with enhanced
carbon.  }
%\end{figure} 

\vskip 0.1in

\noindent
of 912~\AA\ redistributes the flux in the extreme
ultraviolet (EUV) to longer wavelengths.  In an atmosphere with
enhanced helium, the opacity below 912~\AA\ is greatly reduced, and so
much more flux is radiated in the EUV at the expense of the flux at
longer wavelengths.  Enhancing the carbon abundance along with the
helium restores some of this EUV opacity, but the resulting spectrum
is still redder and fainter in the far-UV than that of a normal
stellar atmosphere.  Replacing carbon with nitrogen
can also provide this EUV opacity.

\section{Nature of Subluminous EHB Stars} \label{secsubhb}

Having explored the theoretical implications of high mass loss on the
RGB, we now return to the observations of NGC~2808 and its unusual HB
morphology.  Given our choice of distance modulus, the BHB stars at
$\rm T_{eff} < 15,000~K$ in our STIS CMD (Figure~\figcmd) fall within
the expected range~~of~~lumi- \\

\vskip 6.88in

\noindent
nosities, while hotter stars begin to fall
below the ZAHB.  Note that this is not simply a translation of the
entire HB population to fainter magnitudes: the EHB population has
approximately twice the expected luminosity width in the far-UV.
Thus, one needs to explain both the faint luminosities of the EHB
stars and the EHB luminosity width in order to understand the STIS
CMD.

As suggested in the preceding sections, the subluminous EHB stars in
NGC~2808 may be the progeny of stars that underwent flash mixing on
the WD cooling curve and that have arrived on the ZAHB with a greatly
modified envelope composition.  We will now examine this possibility
in more detail, by translating the blue-hook models in \S\ref{secbh}
and \S\ref{secbhh} into the STIS CMD using the stellar atmospheres in
\S\ref{secbhspec}.  Before doing this, however, we will first rule out
several alternative explanations involving photometric errors, larger
reddening, and radiative levitation.

%\begin{figure}
%\plotfiddle{fig12.eps}{3.0in}{0}{100}{100}{-250}{0}
%\caption{
\parbox{3.25in}{\epsfxsize=3.25in \epsfbox{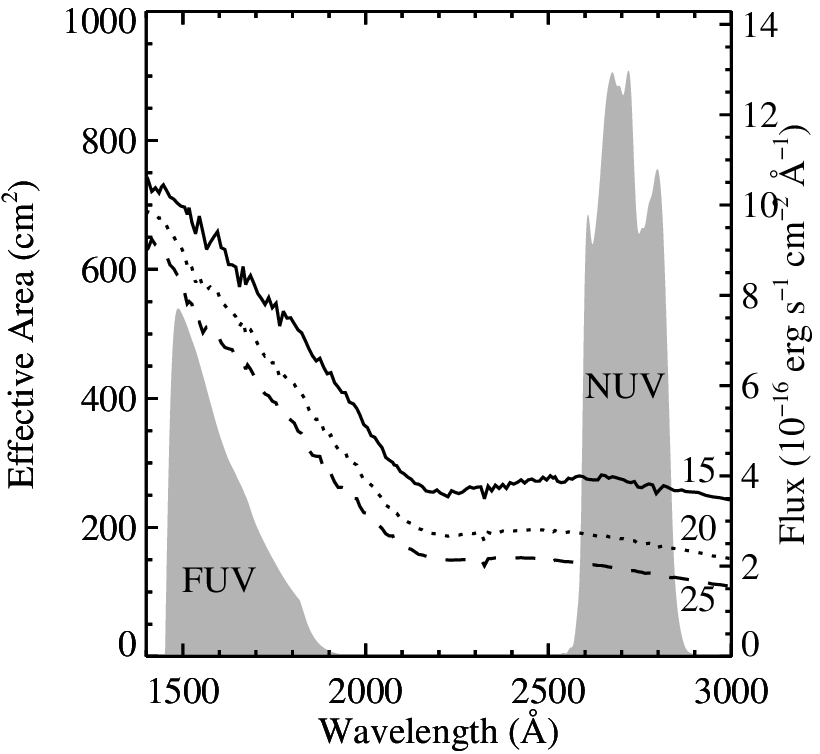}} \\

\parbox{3.25in}{\small {\sc Fig.~\figbandtwo--}
Spectral energy distributions for stars at three different
temperatures along the ZAHB: 15,000~K, 20,000~K, and 25,000~K (labeled).
When compared to the STIS bandpasses (shaded curves), it is apparent
that no calibration error could spread the EHB stars to fainter
magnitudes, while not doing the same to the BHB stars.  Also,
the EHB stars should be as well-detected as the BHB stars, given the
similar flux levels.
}
%\end{figure} 

\medskip

\subsection{Alternative Explanations}

\subsubsection{Photometric Errors} 

The increased luminosity width on the EHB cannot be due to statistical
errors in the photometry, because these errors are very small.  We are
not underestimating the size of these statistical errors on the EHB,
because the luminosity width of the stars on the BHB falls within the
expectations, and these stars are at approximately the same far-UV
luminosity as the canonical EHB.  One would not expect a change of one
magnitude in the near-UV bandpass to dramatically increase the
statistical scatter from $< 0.1$~mag to $\sim$1~mag.  Note that in
these bandpasses, EHB spectra are not very different from BHB spectra
(see Figure~\figbandtwo).  In this figure, we show the theoretical
spectra of three stars lying at different temperatures along the ZAHB.
Because the shape and intensity of each spectrum is similar to the
others, it is apparent that the statistical errors should not vary
dramatically as one moves from $\rm T_{eff} = 15,000~K$ to 25,000~K.

By similar reasoning, the increased luminosity width on the EHB
cannot be due to systematic errors in the data.  Both the EHB stars
and the BHB stars are spread over the entire STIS image, and thus
calibration problems (e.g., flat fielding errors, geometric distortion
errors, focus changes, variability in exposure depth, hot pixels, dead
pixels) should affect both classes of stars similarly.  Furthermore,
there is no change that can be made to the assumed sensitivity curves
that would only depress the far-UV luminosity of some EHB stars
while leaving other EHB stars and all BHB stars unchanged (see
Figure~\figbandtwo), because these stars have similar spectra.
In the same manner, no change in the assumed reddening law could
depress some of the EHB stars relative to the other EHB stars and
BHB stars.

\subsubsection{Reddening}

One might imagine that significantly increasing the assumed extinction
could cause the theoretical EHB locus to drop off more sharply as a
function of increasing temperature, which in turn might increase the
agreement with the STIS CMD, but this is not the case.  As mentioned
earlier, the foreground reddening and distance toward NGC~2808 are
somewhat uncertain.  Ferraro et al.\ (1990\markcite{F90}) compiled a
list of $E(B-V)$ determinations for NGC~2808 from the literature, with
the highest being well above the value reported in most studies:
$E(B-V) =0.34$~mag (Burstein \& McDonald 1975\markcite{BM75}).  The
spectra in Figure~\figbandtwo\ demonstrate that it is difficult to
depress the EHB relative to the BHB in the STIS UV bandpasses, because
the EHB and BHB spectra have similar shape and
intensity. Nevertheless, we demonstrate the effect of increased
reddening in Figure~\figext.  In this figure, we increased $E(B-V)$ to
0.34~mag while decreasing the distance, such that the BHB would still
fall mostly within the theoretical HB locus.  It is obvious that no
choice of distance at this reddening will give agreement across the
entire HB, and that increasing the reddening does not turn the HB
significantly downward on the hot end.  Moreover, the hottest EHB
stars in Figure~\figext\ lie blueward of the canonical EHB, and are
thus unexplained.

Both Walker (1999\markcite{W99}) and Bedin et al.\
(2000\markcite{BPZ00}) bring up the possibility of differential reddening
toward this cluster, of $\sim 0.02$ mag.  Differential reddening could
increase the scatter in the STIS CMD, but then we would see increased
scatter for both the BHB and EHB, not just the EHB.  Furthermore, the
change in $m_{FUV}$ and $m_{NUV}$ is only $\sim 0.15$~mag for a
change in $E(B-V)$ of 0.02~mag, so it would take a very large
differential reddening to reproduce the increased luminosity width
seen on the EHB.  Finally, the STIS field is much smaller than the
ground-based fields, and so any differential reddening should be much
smaller in the STIS photometry than in the ground-based
photometry.

\subsubsection{Radiative Levitation} \label{secradlev}

Grundahl et al.\ (1999\markcite{G99}) have found photometric evidence
that atmospheric diffusion plays an important role in HB morphology.
That such processes may help explain HB anomalies was first suggested
by Caloi (1999\markcite{C99}); later, spectroscopy of BHB stars by
Behr et al.\ (1999\markcite{BCM99}), Behr, Cohen, \& McCarthy
(2000\markcite{BCM00}), and Moehler et al.\ (2000\markcite{M00})
confirmed that atmospheric diffusion can strongly affect BHB star
abundances.  In their CMDs of Galactic globular clusters with extended
horizontal branches, Grundahl et al.\ (1999\markcite{G99}) found that
BHB stars hotter than $\sim 11,500$~K are brighter in Str$\ddot{\rm
o}$mgren $u$ than predicted by canonical HB models (using cluster
abundances), while the EHB and RHB stars agree with the models.  They
explain this jump in the $u$ magnitude by invoking radiative
levitation of heavy elements in stars spanning the range
$11,500~\lesssim \rm T_{eff} \lesssim $~20,000~K.  This levitation
greatly enhances the abundance of metals in the stellar atmosphere.
The Grundahl et al.\ (1999\markcite{G99}) study showed that the effect
on the CMD was strongly dependent upon the bandpasses used.  At high
atmospheric metallicity (up to [Fe/H]=+1.0), BHB stars become brighter
(by $\lesssim$~0.3 mag) in Johnson $U$ and Str$\ddot{\rm o}$mgren $u$,
but become somewhat fainter (by $\lesssim 0.1$~mag) in WFPC2/F160BW
and UIT/1620~\AA.  Subsequent observations by Bedin et al.\
(2000\markcite{BPZ00}) demonstrated that the same effect is present in
the BHB of NGC~2808.

Our STIS far-UV bandpass is similar in wavelength coverage to the
WFPC2/F160BW bandpass studied by Grundahl et al.\
(1999\markcite{G99}), although it is far more sensitive.  If
atmospheric diffusion is affecting the abundances of the BHB stars
only, one would expect a small depression of the BHB luminosity
relative to that of the EHB and RHB stars, when in fact we see the
opposite: a depression~~of~~the~~EHB~~luminosity~~relative to the 

%\begin{figure}
%\plotfiddle{fig13.eps}{6.0in}{0}{100}{100}{-250}{0}
%\caption{
\hskip 0.25in
\parbox{6.5in}{\epsfxsize=6.5in \epsfbox{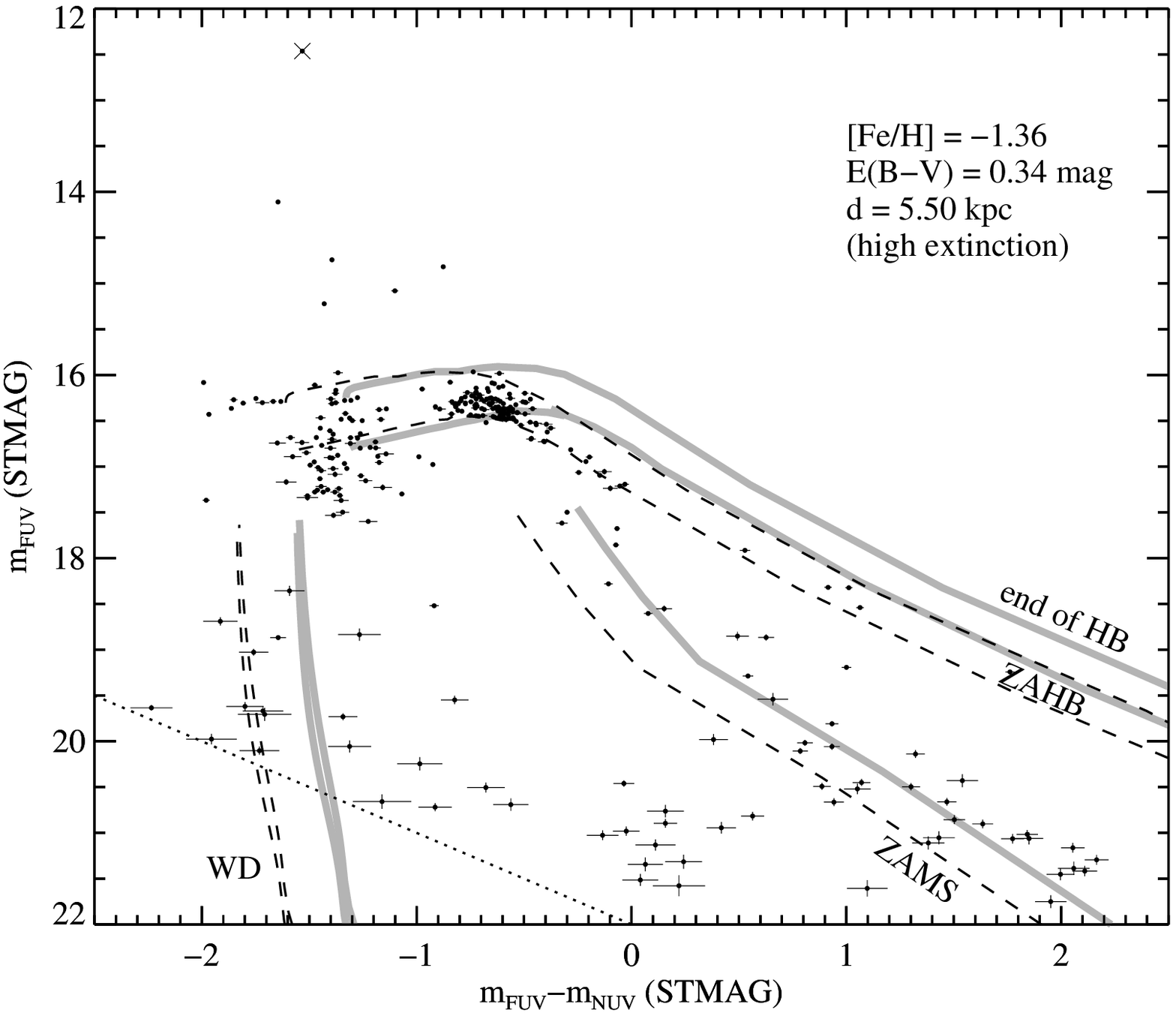}}

\vskip -0.1in
\hskip 0.25in
\parbox{6.5in}{\small {\sc Fig.~\figext--}
Theoretical loci (grey curves) for the HB, ZAMS, and WD evolutionary
phases in the same STIS CMD as shown in Figure~\figcmd.  We translated
the theoretical models to the observational plane while assuming
a larger foreground reddening of $E(B-V)=0.34$~mag.  The distance
modulus was set by matching the theoretical ZAHB to the lower boundary
of the BHB stars.  The dashed curves show the location of the theoretical
loci when the nominal cluster parameters are assumed, as in Figure~\figcmd.
This figure demonstrates that an increase in the assumed extinction
cannot depress the EHB relative to the BHB.
}
%\end{figure} 

\medskip

\noindent
BHB
luminosity.  Thus the Grundahl et al.\ (1999\markcite{G99}) results
suggest that atmospheric diffusion does not explain the discrepancy
seen in the STIS CMD of NGC~2808.  Nonetheless, it is worth exploring
the effect of enhanced atmospheric metallicity in our CMD.  If, for
some reason, our far-UV bandpass was to respond in the same manner as
Str$\ddot{\rm o}$mgren $u$ (which lies at longer wavelengths than the
STIS far-UV bandpass), then the BHB stars would be brighter than
expected, which would give the misleading impression that the EHB was
too faint compared to the BHB.  However, this would still not explain
the unusually large EHB luminosity width.

To demonstrate the effect of atmospheric diffusion on our CMD, we
translate the entire HB (regardless of $\rm T_{eff}$) to the
observational plane using synthetic spectra at [Fe/H]=+1.0 in
Figure~\figdif.  The use of these models is meant to show the maximum
effect of radiative levitation; our large metallicity enhancement is
much higher than what is spectroscopically observed in BHB stars by
Behr et al.\ (1999\markcite{BCM99}; 2000\markcite{BCM00}), in M13 and
M15, and Moehler et al.\ (2000\markcite{M00}), in NGC~6752, where
[Fe/H] is \\

\vskip 6.58in

\noindent
generally solar to a few times solar.  For the range of
temperature spanned by the BHB stars, a theoretical HB locus at solar
metallicity (not shown in Figure~\figdif) is very nearly coincident
with the HB locus at the cluster metallicity (dashed), and thus the
BHB stars in the STIS data appear at the expected luminosity for the
likely range of atmospheric diffusion effects.  Note that our use of
scaled-solar models provides only an approximation to the effects of
radiative levitation, because the light-element abundances generally
do not show the strong abundance enhancements seen for iron.
Figure~\figdif\ demonstrates that in the STIS bandpasses, HB stars
become both fainter and redder at high atmospheric metallicity,
regardless of $\rm T_{eff}$; i.e., the metallicity affects the far-UV
band more strongly, so the translation is not simply downward in the
CMD.  It is clear from the figure that the BHB stars will not increase
in luminosity if they are affected by atmospheric diffusion; they will
decrease in far-UV luminosity, as expected.

What if diffusion is affecting the EHB stars more strongly than the
BHB stars?  In that case, the EHB stars would become 

%\begin{figure}
%\plotfiddle{fig14.eps}{6.0in}{0}{100}{100}{-250}{0}
%\caption{
\hskip 0.25in
\parbox{6.5in}{\epsfxsize=6.5in \epsfbox{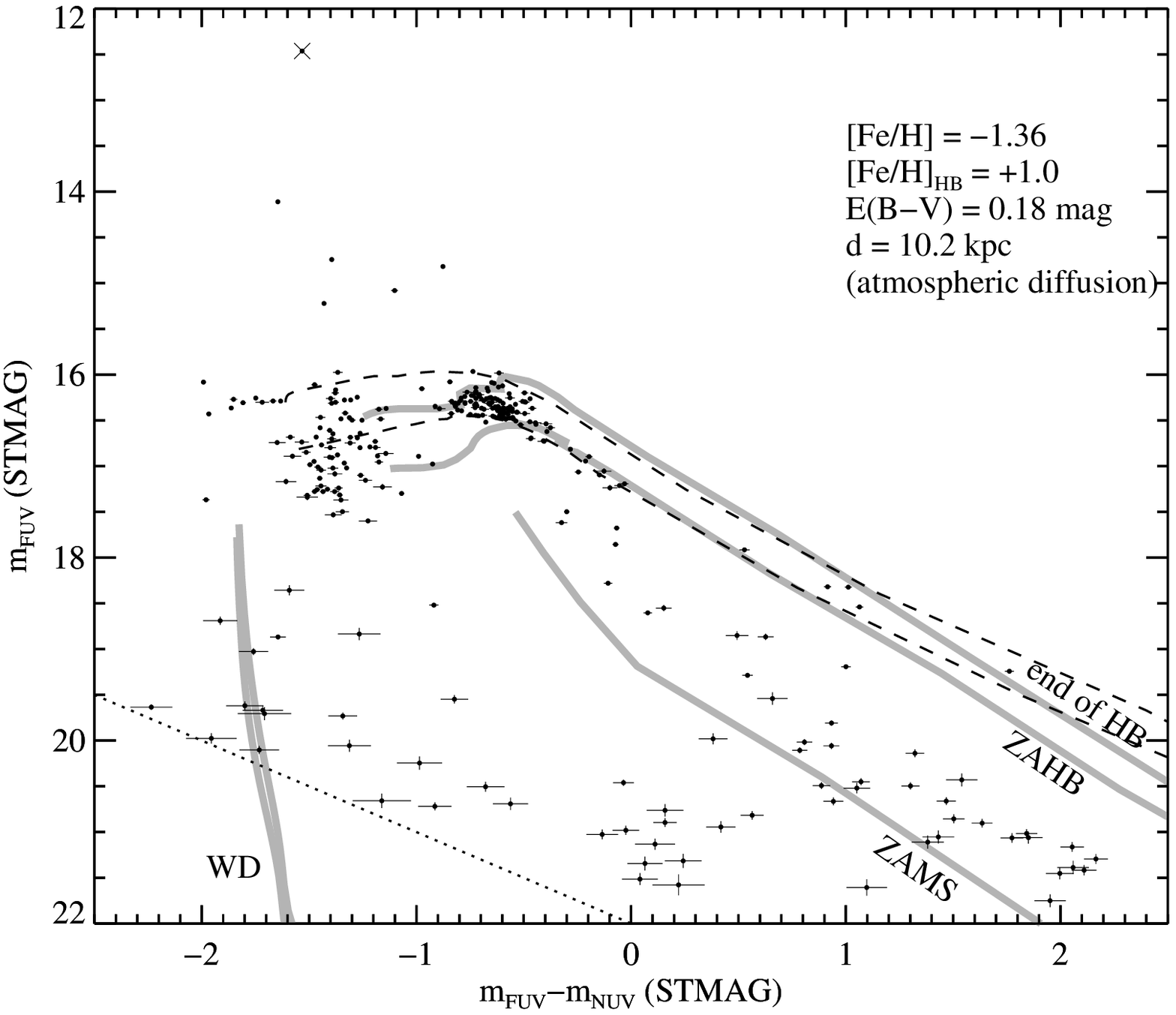}} 

\vskip -0.1in
\hskip 0.25in
\parbox{6.5in}{\small {\sc Fig.~\figdif--}
The same STIS CMD shown in Figure~\figcmd, but with the
entire theoretical HB translated to the observational plane (grey
curves) while assuming an atmospheric abundance of [Fe/H]=+1.0 (note
that atmospheric diffusion is only thought to affect HB stars at $\rm
11,500 \lesssim T_{eff} \lesssim 20,000~K$).  This
metallicity-enhanced HB locus demonstrates that atmospheric diffusion
cannot sufficiently depress the EHB relative to the BHB or increase
the luminosity width of the EHB.  The dashed curves show the location
of the theoretical loci when the nominal cluster parameters are
assumed, as in Figure~\figcmd.  }
%\end{figure} 

\medskip

\noindent
fainter, relative
to the BHB, but this would contradict the Grundahl et al.\
(1999\markcite{G99}) results, which show that the photometric effects
of diffusion seem to be decreasing near $\rm T_{eff} \sim 20,000~K$.
Furthermore, the EHB stars in the STIS CMD are not completely
translated to fainter magnitudes; they are spread to fainter
magnitudes, with an increased luminosity width.  The only way to
increase the far-UV luminosity width on the EHB is to assume a
variable metallicity enhancement for the EHB stars.  However, if one
does this, the EHB stars in the data would still have a luminosity
width that is larger than this ad hoc scenario.  There is also no
obvious way to account for the hottest EHB stars that lie blueward of
the canonical EHB locus in Figure~\figdif.  In summary, atmospheric
diffusion does not appear to explain the discrepancies between the
data and the models.

\subsection{Blue-hook Explanation} \label{secbhexp}

Blue-hook stars were the final explanation that we considered for the
subluminous EHB stars in our CMD.  These stars would coexist with~~the
canonical EHB stars, instead~~of~~replac- \\

\vskip 6.53in

\noindent
ing them, so they could
potentially widen the apparent luminosity width of the EHB.  To
demonstrate where blue-hook stars would lie in the STIS CMD, we
translated the ZAHB model of each $\eta_R = 0.860$ blue-hook sequence
(see Figure~\figbhevol) to the observational plane.  As evident in
Figure~\figbhevol, the exact choice of $\eta_R$ makes little
difference, because all values of $\eta_R$ within a composition set
($0.818 \leq \eta_R \leq 0.936$) have very similar tracks; we chose
$\eta_R = 0.860$ simply because it is near the middle of the range.

Figure~\fighook\ shows the location of these ($\eta_R =0.860$)
blue-hook models in the STIS CMD.  The circle shows the ZAHB model in
the ``H envelope'' blue-hook track (see Figure~\figbhevol\ and
\S\ref{secbh}), translated to the observational plane using a model
atmosphere and synthetic spectrum with the nominal cluster abundance.
Because the spectrum is dominated by hydrogen opacity, it is marked
with an ``H.''  The square and diamond show, respectively, the ZAHB
models for the ``He envelope'' and ``H+C envelope'' blue-hook tracks,
translated to the observational plane with the appropriate atmospheres
and spectra.  The grey curve 

%\begin{figure}
%\plotfiddle{fig15.eps}{6.0in}{0}{100}{100}{-250}{40} \vskip -0.75in
%\caption{
\hskip 0.25in
\parbox{6.5in}{\epsfxsize=6.5in \epsfbox{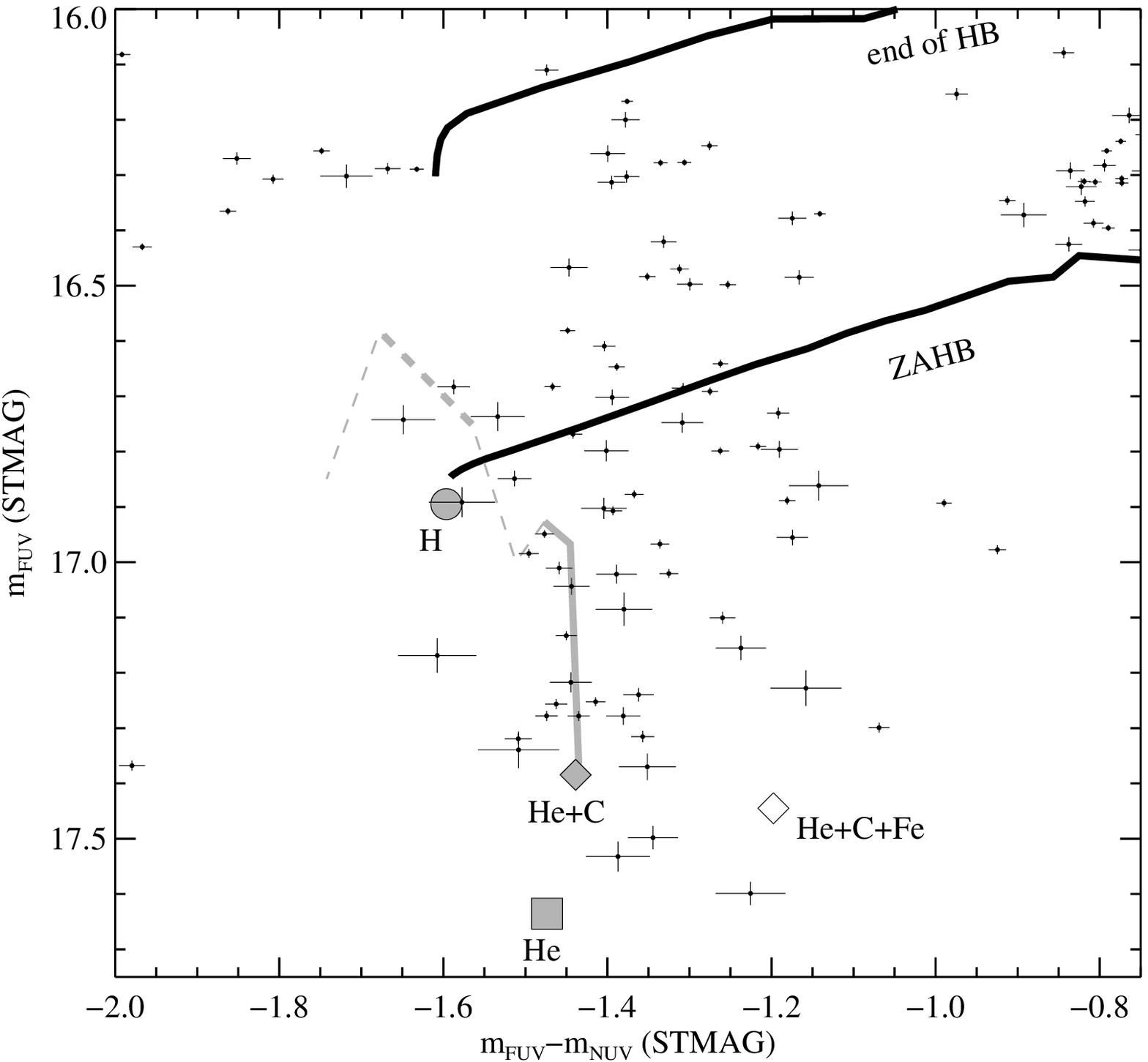}} 

\vskip 0.1in
\hskip 0.25in
\parbox{6.5in}{\small {\sc Fig.~\fighook--}
Location of the $\eta_R=0.860$ blue-hook ZAHB models in the
STIS CMD.  Blue-hook stars with normal H envelopes (circle) would
begin their core He-burning evolution near the hot end of the canonical EHB,
and thus do not explain the stars below the canonical ZAHB (labeled).
Blue-hook stars with enhanced helium (square) or enhanced helium and
carbon (diamond) atmospheres would begin their core He-burning evolution
well below the canonical ZAHB, and then evolve to brighter
luminosities (grey curve), filling in the area under the canonical ZAHB.
If radiative levitation in the hot He+C models increased the iron
abundance to [Fe/H]=+1.0, the models would become significantly redder
(open diamond).  The dashed grey curve shows the post-HB evolution of
the He+C model, with thicker dashes denoting the slowest phase.}
%\end{figure} 

\medskip

\noindent
shows the HB evolution of the ``He+C''
model, with the post-HB evolution shown as a dashed line.

Figure~\fighook\ demonstrates that blue-hook stars with normal H
envelopes cannot explain the subluminous stars in the STIS CMD: they
are predicted to lie near the hot end of the canonical ZAHB.  Looking
at Figure~\figbhevol, this is not surprising, because the H blue-hook
sequences also lie near the canonical ZAHB in physical parameter
space.  The small $\sim 0.1$~mag drop in the far-UV luminosity comes
from the small decrease in core mass, as discussed in \S\ref{secevol}
and also shown in D'Cruz et al.\ (1996\markcite{D96},
2000\markcite{D00}).  In contrast, the He and He+C models in
Figure~\fighook\ show the same dramatic drop in far-UV luminosity that
is present in the STIS data.  Roughly half of the luminosity drop seen
in the He and He+C models comes from the larger bolometric correction
associated with their higher temperatures, as compared to the H
blue-hook model; the rest of the luminosity drop and all of \\

\vskip 7.04in

\noindent
the
movement to the red comes from the effects of the envelope abundance
changes on the emergent spectra.  Thus, the location of blue-hook
models in a CMD is very sensitive to the composition of the envelope
and atmosphere, but not very sensitive to the reduced core mass.  It
is worth noting here that the distribution of metallicity in
$\omega$~Cen might drive differences in its blue-hook morphology.

The blue-hook stars will evolve through their core He-burning phase in
much the same way as canonical HB stars (see Figure~\figbhevol): they
will slowly evolve $\sim$0.5~mag brighter in bolometric luminosity at
roughly constant $\rm T_{eff}$.  They will then evolve more rapidly to
hotter $\rm T_{eff}$ and brighter luminosities, and go through a phase
very similar to the canonical AGB-Manqu$\acute{\rm e}$ evolution.
Thus, the spread of stars in the STIS CMD between the blue-hook ZAHB
(diamond) and the canonical ZAHB is due to the evolution of blue-hook
stars to higher luminosities (grey curve in Figure~\fighook) as they
undergo core He-burning.  Together, the blue-hook and canonical stars
would predict an EHB luminosity width very close to that observed in
the STIS CMD.
 
Changes in the assumed atmospheric abundances could produce some
scatter in the colors of the blue-hook stars.  In \S\ref{secbh}, we
found that the envelope of blue-hook stars would likely have greatly
enhanced helium and carbon, and this corresponds to the ``He+C'' point
in Figure~\fighook.  If the carbon abundance is not as high as 4\% by
mass, a blue-hook star will lie somewhere between the ``He+C'' point
and ``He'' point.  Although the supply of protons in the stellar
envelope during flash mixing is not large enough to burn the carbon to
nitrogen, replacing the enhanced carbon by enhanced nitrogen would
move the He+C point (diamond) $\sim$0.05~mag to the blue in
Figure~\fighook.

The colors of the blue-hook stars would also be affected by the
radiative levitation of iron in the stellar atmosphere.  For example,
if the surface abundance of iron increased to [Fe/H]=+1.0, the
$m_{FUV}-m_{NUV}$ color of a star at the same $\rm T_{eff}$ would be
$\sim$0.25~mag redder than the He+C point.  This is shown by the open
diamond in Figure~\fighook, marked ``He+C+Fe.''  An enhanced iron
abundance at the surface might also increase the stellar radius
slightly, thus decreasing the $\rm T_{eff}$, and this would also move
the He+C point somewhat redder.  The iron enhancement seen in BHB
stars does not appear to be present in the cooler EHB stars (see
\S\ref{secradlev}; also see Bedin et al.\ 2000\markcite{BPZ00}), but
diffusion calculations (Charpinet et al.\ 1997\markcite{C97}) show
that the surface iron abundance can be enhanced to solar or even
super-solar abundances at $\rm T_{eff} \gtrsim$~30,000~K, which is
hotter than the canonical EHB but appropriate for blue-hook stars;
these iron enhancements offer a possible driving mechanism for the
pulsating sdB stars, which show solar iron abundances (Heber, Reid, \&
Werner 2000\markcite{H00}).  Furthermore, it should be easier to
enhance the iron abundance through radiative levitation if the
atmosphere is mostly helium instead of hydrogen, because of the
decrease in EUV opacity and the increase in the mean molecular weight
in the atmosphere.

The blue-hook explanation of the subluminous EHB stars can be tested
by comparing the location of the blue-hook models in different
bandpasses with the observational data.  For example, NGC~2808 was
observed by Sosin et al.\ (1997\markcite{SDD97}) with WFPC2, using the
F218W, F439W, and F555W bands.  An examination of their figures shows
a significant number of stars below the ZAHB, but the luminosity
offset is somewhat smaller than in our STIS CMD.  However, the F439W
and F218W filters are not as sensitive to the anomalous abundances in
flash-mixed blue-hook stars, as one might expect from an examination
of Figure~\figatmos; as one moves from the far-UV to longer
wavelengths, the He+C spectrum moves closer to the spectrum from a
canonical EHB star.  We further demonstrate this point in
Figure~\figopt, which shows the translation of the canonical ZAHB and
blue-hook ZAHB models to the STIS bandpasses, the WFPC2 bandpasses,
and ground-based bandpasses.  Each panel in Figure~\figopt\ has the
same range in color (1~mag) and luminosity (2~mags).

At a given color, the luminosity offset for the blue-hook models is
smaller in the WFPC2/F439W bandpass than in the STIS/FUV bandpass.
Also, in the WFPC2 CMD, He+C blue-hook stars should extend bluer than
the hot end of the canonical ZAHB, while in the STIS CMD, these stars
are redder than the hot end of the ZAHB.  Thus, in the WFPC2
bandpasses, the He+C blue-hook stars would appear to lie near an
extension of the canonical ZAHB, if it was drawn to arbitrarily small
envelope masses and high $\rm T_{eff}$.  In fact, the theoretical ZAHB
employed by Sosin et al.\ (1997\markcite{SDD97}) was extended to
$M_{env} = 10^{-4} M_{\odot}$ and $\rm T_{eff} \approx$~40,000~K,
which is far hotter than the true termination of the canonical ZAHB
(see \S\ref{sechb} and Figure~\figeta).  Note also that Sosin et al.\
(1997\markcite{SDD97}) transformed $m_{F439W}$ to an approximate
Johnson $B$, while we have retained the STMAG system for the WFPC2
bandpasses; for a synthetic spectrum at $\rm T_{eff} = 25,000$~K,
log~$g$~=5, and [Fe/H]~$=-1.5$, $B = m_{F439W} + 0.66$~mag.  Thus our
models would appear to be approximately 0.7~mag fainter in $B$ than in
$m_{F439W}$.  The predicted $B$ magnitude of the He+C blue model in
Figure~\figopt $b$ agrees well with the faint end of the EHB in the
Sosin et al.\ (1997\markcite{SDD97}) ($m_{F218W} - B, B$) CMD.

Bedin et al.\ (2000\markcite{BPZ00}) show a ($U-B, U$) CMD of
NGC~2808 in which the EHB stars are well-matched by an extension of
the canonical ZAHB to very small envelope masses ($M_{env} = 4 \times
10^{-4} M_\odot$).  In particular, there is no indication of
subluminous stars in this CMD.  This again demonstrates that the
unusual nature of the blue-hook stars is most easily discerned in UV
bandpasses.  Comparison of the Bedin et al.\ (2000\markcite{BPZ00})
CMD with Figure~\figopt $c$ shows excellent agreement: the blue-hook
stars are predicted to lie directly along an extension of the hot EHB,
just as observed.  There is, however, a clear discrepancy between the
faint end of the EHB at $U \sim 20.0$~mag in the Bedin et al.\
(2000\markcite{BPZ00}) CMD and the faint end of the canonical ZAHB at
$U \sim 19.4$~mag in Figure~\figopt $c$.  The fainter observed limit
for the EHB is entirely consistent with the predicted $U$ magnitudes
of the blue-hook models.  We note that this discrepancy is less
pronounced in the Bedin et al.\ (2000\markcite{BPZ00}) CMD because the
canonical ZAHB plotted in this CMD extends to smaller $M_{env}$ (and hence
higher $\rm T_{eff}$) than one would expect from the present calculations or
the calculations of D'Cruz et al.\ (1996\markcite{D96}).

     The ($B-V$,$V$) CMD in Figure~\figopt $d$ also shows that the
canonical ZAHB and the flash-mixed blue-hook models predict different
limiting magnitudes for the faint end of the EHB.  The canonical HB
ends at $V \sim 20.4$~mag, while the blue-hook models extend another
$\sim 0.5$~mag fainter to $V\sim 20.9$~mag.  The deep photometry of
Walker (1999\markcite{W99}) and Bedin et al.\ (2000\markcite{BPZ00})
shows that the faint end of the blue HB tail in NGC~2808 is located at
$V\sim 21.2$~mag, in good agreement with the predicted location of the
blue-hook models, but not with the faint end of the canonical ZAHB.
As noted in \S\ref{secevol}, there is no way within the canonical
framework to extend the canonical ZAHB to fainter magnitudes.  It
appears, therefore, that canonical models cannot account for the
extent of the blue HB tail in NGC~2808 in either the ($U-B$,$U$) or
($B-V$,$V$) CMDs.

    In summary, the flash-mixed blue-hook models, when translated to
the observational plane, are able to explain the luminosities of the
faintest subluminous EHB stars, as well as the large luminosity width
of the EHB in NGC~2808.  Moreover, they are consistent with the
locations of the hottest EHB stars in various observational
bandpasses.

\section{EHB GAP} \label{secgap}

There are three significant gaps in the NGC~2808 HB distribution.  The
gap between the BHB and RHB stars is stretched in the STIS bandpasses,
and thus there are very few stars on the HB at $m_{FUV}-m_{NUV} >
0$~mag in Figure~\figcmd.  At present, there is no plausible
explanation for this strong bimodality of the HB distribution in
NGC~2808.  Sosin et al.\ (1997\markcite{SDD97}) reported two more gaps
in the NGC~2808 HB distribution: one between the EHB and BHB stars,
and one within the EHB itself.  These gaps were confirmed in optical
CMDs by Walker (1999\markcite{W99}) and Bedin et
al. (2000\markcite{BPZ00}).  Our STIS CMD clearly shows the EHB-BHB
gap, but the gap within the EHB is not present.

%\begin{figure}
%\plotfiddle{fig16.eps}{6.0in}{0}{100}{100}{-250}{45}
%\vskip -0.9in
%\caption{
\hskip 0.25in
\parbox{6.5in}{\epsfxsize=6.5in \epsfbox{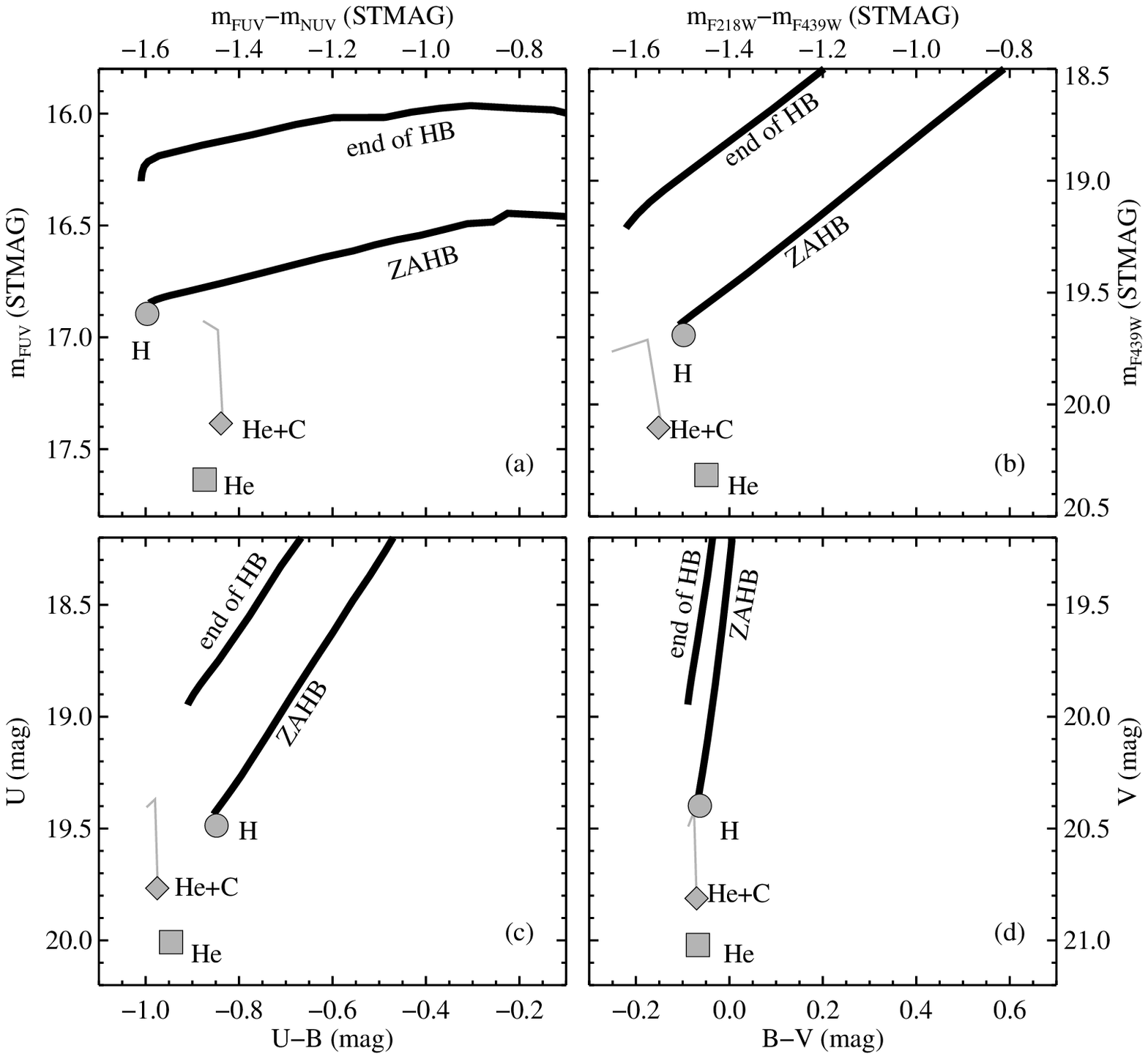}} \\ 

\hskip 0.25in
\parbox{6.5in}{\small {\sc Fig.~\figopt--}
Locations of the blue-hook ZAHB models and the canonical HB
in four sets of bandpasses.  The HB evolution of a blue-hook He+C
model is shown as a grey curve, as in Figure~\fighook. The stretch in
color and luminosity is the same in each panel.  Upper left: In the
STIS bandpasses ($m_{FUV}-m_{NUV}$,$m_{FUV}$), the He+C blue-hook
model lies fainter and redder than the hot end of the canonical ZAHB.
Upper right: In the WFPC2 bandpasses
($m_{F218W}-m_{F439W}$,$m_{F439W}$), the He+C blue-hook model appears
fainter and bluer than the end of the canonical ZAHB.  Lower left: In
a ground-based ($U-B$,$U$) CMD, the He+C model appears to lie along an
extension of the canonical ZAHB.  Lower right: in a ground-based
($B-V$,$V$) CMD, the EHB is nearly vertical, and again the He+C model
appears to lie along an extension of the canonical ZAHB.  }
%\end{figure} 

\medskip

As noted in section \S\ref{secflash}, flash mixing should be a normal
consequence of helium ignition on the WD cooling curve.  In this
section we will argue that the dichotomy produced by such mixing
between the canonical EHB and blue-hook models is responsible for the
EHB gap in the optical CMDs of NGC 2808.  However, this possibility
raises an important question: why are EHB gaps not apparent in the
optical CMDs of other globular clusters with extended blue HB tails?
For example, the optical CMDs of $\omega$~Cen reported by Kaluzny et
al.\ (1997\markcite{KKS97}) and Lee et al.\ (1999\markcite{LJS99}) do
not seem to show an EHB gap despite the substantial population of EHB
stars.  Quite possibly, any EHB gap in $\omega$~Cen has been blurred
by the metallicity distribution in the cluster.  Another candidate for
an EHB gap is NGC~6273, which has one of the longest blue HB tails of
any globular cluster (Piotto et al.\ 1999\markcite{P99}).
Unfortunately, the CMD of NGC~6273 is affected by large differential
reddening ($\Delta E(B-V) \sim 0.2$~mag; Piotto et al.\
1999\markcite{P99}), which would likewise obscure any EHB gap.  Other \\

\vskip 6.88in

\noindent
globular clusters with long blue HB tails, e.g., M13, M80 and
NGC~6752, do not contain a sufficient number of the faintest EHB stars
($M_V \gtrsim 4.5$~mag) to determine if an EHB gap is present (see,
e.g., Figure 7 of Piotto et al.\ 1999\markcite{P99}).  Thus NGC~2808
stands out as the globular cluster where an EHB gap is most easily
detected observationally.

The absence of the EHB gap in the STIS CMD can be understood if one
examines Figure~\fighook.  In the STIS bandpasses, the blue-hook
models begin their core He-burning evolution redder and fainter than
the canonical EHB, and then evolve to brighter luminosities (along the
grey curve), filling in the area under the canonical ZAHB with
``subluminous'' HB stars.  Thus one would not predict a gap in the
STIS CMD once the HB evolution is taken into account.  However, in the
WFPC2 and ground-based bandpasses (Figure~\figopt), the blue-hook
models lie along an extension of the canonical ZAHB, separated by a
gap in luminosity.

%\begin{figure}
%\plotfiddle{fig17.eps}{6.0in}{0}{100}{100}{-250}{45}
%\vskip -0.9in
%\caption{
\hskip 0.25in
\parbox{6.5in}{\epsfxsize=6.5in \epsfbox{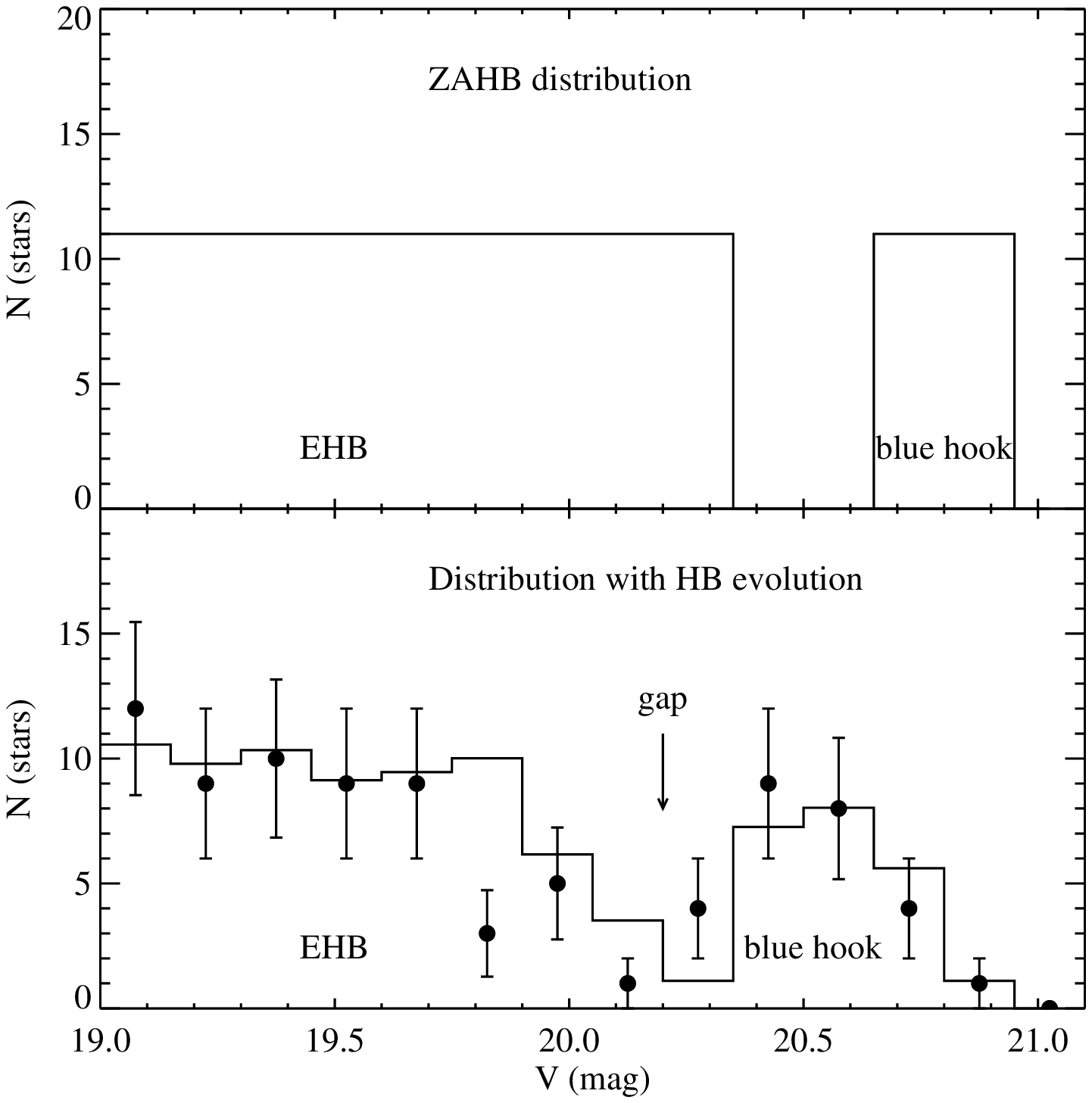}} 

\vskip -0.1in

\hskip 0.25in
\parbox{6.5in}{\small {\sc Fig.~\figgap--}
Top panel: Assumed uniform distribution in $V$ along the
ZAHB, with a gap at $V \approx 20.5$~mag between the EHB and the He+C
blue-hook stars (compare to Figure~\figopt $d$).  Bottom panel: Predicted
distribution with HB evolution from the ZAHB distribution in the top
panel.  Note that the gap has shifted closer to $V \approx 20$~mag.
For comparison, we plot the EHB luminosity function from Bedin et al.\
(2000\protect\markcite{BPZ00}) as points with their associated Poisson
errors.  The observed gap within the EHB coincides closely with the
predicted gap between the EHB and blue-hook stars.  }
%\end{figure} 

\medskip

The EHB gap is very obvious in the ($B-V$,$V$) CMD of NGC~2808 of
Bedin et al.\ (2000\markcite{BPZ00}) at $V \approx 20$~mag.  Because
the EHB is almost vertical in a ($B-V$,$V$) CMD, they were able to
use the luminosity function in $V$ to analyze this gap,
concluding that this gap is probably real and not a statistical
fluctuation.  The separation between the blue-hook models and the
canonical EHB, shown in Figure~\figopt $d$, demonstrates why one
expects a hot gap in the ($B-V$,$V$) plane.  However, the gap in
Figure~\figopt $d$ refers only to ZAHB models.  To understand where
the gap would actually be observed in a CMD, we must include the HB
evolution of the blue-hook and EHB models to brighter $V$ magnitudes.

    We show in Figure~\figgap\ the predicted location of the gap
between the canonical EHB and blue-hook models, compared to the Bedin
et al.\ (2000\markcite{BPZ00}) luminosity function (shown as filled
points with error bars).  To derive the theoretical luminosity
function in Figure~\figgap $b$, we first assumed that the EHB and
blue-hook models were distributed uniformly in $V$ along the ZAHB
(see Figure~\figgap $a$).  In addition, we adopted a bin size of \\

\vskip 6.58in

\noindent
0.15~mag in $V$ to be consistent with the Bedin et al.\
(2000\markcite{BPZ00}) analysis.  The total number of stars in the
ZAHB distribution was normalized so that the theoretical HB
distribution in Figure~\figgap $a$ had approximately the same number
of stars as the Bedin et al.\ (2000\markcite{BPZ00}) luminosity
function from $19 \leq V \leq 21$~mag.  The $V$ magnitudes of the
blue-hook models in Figure~\figgap $a$ were determined from the ZAHB
luminosities of the He+C blue-hook models with $0.818 \leq \eta_R \leq
0.936$.  The gap in the ZAHB distribution at $V\sim 20.5$~mag
corresponds to the gap between the canonical ZAHB and He+C blue-hook
models in Figure~\figopt $d$.

    The ZAHB models in Figure~\figgap $a$ will evolve towards higher
luminosities until, by the end of the HB phase, they are $\sim 0.5$~mag
brighter in $V$ than at the ZAHB.  In order to include this HB
evolution, we distributed each of the ZAHB models in Figure~\figgap $a$
uniformly in time over its evolutionary track.  These evolved HB
models were then added to the appropriate luminosity bin to obtain the
theoretical HB distribution shown in Figure~\figgap $b$.
          
    Our theoretical luminosity function agrees well with the Bedin et
al.\ (2000\markcite{BPZ00}) data, given the unknown mass distribution
on the HB and the uncertainties in translating the theoretical models
to the observational plane.  In particular, we note the good agreement
between the predicted and observed location of the EHB gap.  These
results support the possibility that the gap within the EHB of
NGC~2808 is due to the dichotomy between the blue-hook and canonical
EHB models.

It is worth stressing that the gap within the EHB will only be
apparent if the blue-hook stars have mixed envelopes.  As shown in
Figure~\figopt $d$, if the blue-hook stars have envelopes with the
normal cluster abundances, the blue-hook stars pile up near the end of
the canonical ZAHB, with no gap between the canonical EHB and
blue-hook stars.  Whitney et al.\ (1998\markcite{WRO98}) argued that
such an effect could explain the gap between the EHB and BHB in
$\omega$~Cen.  They only considered blue-hook stars with normal
atmospheres, and showed that the range of $\eta_R$ giving rise to the
blue-hook stars was larger than the range of $\eta_R$ populating the
HB between the EHB and BHB; this effect would be exaggerated for the
more metal-rich component of the $\omega$~Cen metallicity
distribution (see D'Cruz et al.\ 1996\markcite{D96}).  Thus, the
region between the EHB and BHB would appear underpopulated, and then
there would be a clump of stars at the hot end of the EHB (the blue-hook
stars).  However, the above analysis suggests that the dichotomy
between the blue-hook and EHB stars causes the hottest gap within the
EHB distribution, not the gap between the EHB and BHB, which therefore
remains unexplained.  From Figure~15 of Bedin et al.\
(2000\markcite{BPZ00}), we see that there are more canonical EHB stars
than blue-hook stars, even though the range in $\eta_R$ that populates
the canonical EHB is small.  The gap between the EHB and BHB, at $V
\approx 18.5$~mag, is too bright to be due to a build up of blue-hook
stars near the hot end of the canonical ZAHB.

We have no plausible explanation for the EHB-BHB gap, which is very
obvious in our STIS CMD, at $m_{FUV}-m_{NUV} \approx -1$~mag.
However, an inspection of Figure~\figeta\ may offer a possible clue
for the EHB-BHB gap.  The HB stars that should fall in this gap are
produced (in our models) by $\eta_R \approx 0.740$.  This value of
$\eta_R$ falls in the transition between stars that flash at the RGB
tip and those stars that flash as they are peeling away from the RGB
(compare panels $b$ and $c$ in Figure~\figeta).

\section{IMPLICATIONS} \label{secimp}

\subsection{The Origin of Field He-sdB and He-sdO Stars}

The analysis presented herein offers a possible explanation for the
subluminous EHB stars seen in the globular clusters NGC~2808 and
$\omega$ Cen.  We have shown that stars evolving with high mass-loss
on the RGB will undergo a late helium-core flash on the WD cooling
curve that leads to convective flash mixing of the envelope.
When these stars begin stable core He-burning, they will do so at
temperatures significantly hotter than the hot end of the canonical
EHB.  

In the Galactic field, hot HB stars are observed spectroscopically as
sdB and sdO stars (see Green, Schmidt, \& Liebert 1986\markcite{G86}
for one version of the subdwarf classification scheme).  Many sdO
stars show enhanced helium, and those that do tend to show enhanced
carbon as well (Lemke et al.\ 1997\markcite{L97}).  Only a small
fraction ($\sim 5$\%) of sdB stars show enhanced helium (Jeffery et al.\
1996\markcite{J96}), but again, those that do also show strong carbon
lines (e.g., Moehler et al.\ 1990\markcite{M90}).  The sdO and sdB
stars that do not have enhanced helium tend to be extremely deficient
in their abundances of helium and most heavy metals.  These
deficiencies are attributed to gravitational settling.

Flash mixing would provide a possible explanation for why some sdB and
sdO stars are helium- and carbon-enhanced, while others are depleted.
Stars at $\rm T_{eff} \lesssim 30,000~K$ tend to be classified as sdB,
and stars at $\rm T_{eff} \gtrsim 40,000~K$ tend to be classified as
sdO, with the range $\rm 30,000 \lesssim T_{eff} \lesssim 40,000~K$
variably classified as sdO, sdB, or sdOB.  Because flash mixing on the
WD cooling curve produces stars that are significantly hotter than the
canonical EHB, one would expect more sdO stars to show helium and
carbon enhancement than sdB stars, as observed.  Note that we have no
evidence for subluminous EHB stars or an EHB gap in the log~$\rm
T_{eff}$ / log~$g$ plane in the Galactic field, nor do we have flash-mixed
models for metal-rich stars; it is too early to say how much of a role
these flash-mixed stars play in the Galactic field population, but
these matters deserve further investigation.

The role of flash mixing in the field population might be partly
obscured by several processes.  If metallicity plays a role in RGB
mass loss, the difference in metallicity between the field population
and the globular cluster population may affect the properties of the
flash-mixed HB models.  For example, the more heterogeneous
compositions of the field population might obscure an EHB gap that
would otherwise be evident in a single metallicity stellar population.
Also, the binary fraction in the field subdwarf population is
unusually high (e.g., Maxted et al.\ 2001\markcite{M01}), which
suggests that binarism might play a role in the formation of the field
subdwarfs.  In contrast, the fraction of EHB stars in NGC~2808
relative to the total HB population does not appear to vary radially
(Walker 1999\markcite{W99}; Bedin et al.\ 2000\markcite{BPZ00}),
suggesting that these EHB stars are the products of single star
evolution.  The same is also true for $\omega$ Cen, which has the
largest known population of EHB stars of any globular cluster (D'Cruz
et al.\ 2000\markcite{D00}).  Such possible differences in origin
might also affect the evidence for flash mixing.  Finally,
gravitational settling of helium and carbon is very likely to obscure
the evidence of flash mixing in some fraction of the hot HB stars,
especially if the envelope should contain a small residual amount of
hydrogen, and this settling might itself be affected by metallicity
and binarism.

\subsection{Pulsating Subdwarfs}

The existence of pulsating sdB stars was predicted theoretically by
Charpinet et al.\ (1996\markcite{CFB96}); subsequent observations by
Kilkenny et al.\ (1997\markcite{K97}) of a binary sdB star,
EC~14026-2647, proved their existence.  More pulsating sdB stars were
soon found (see O'Donoghue et al.\ 1999\markcite{O99} and references
therein), and the class is now referred to as EC~14026 stars or sdBV
stars.  Spectroscopy of these stars shows solar iron abundance even
though other metals are greatly depleted (Heber et al.\
2000\markcite{H00}).  The enhancement of iron relative to other metals
is of particular interest, because iron opacity is thought to be the
pulsation driving mechanism (Charpinet et al.\ 1997\markcite{C97}).

Because the pulsating sdB stars tend to lie near $\rm T_{eff} \sim
35,000$~K (O'Donoghue et al.\ 1999\markcite{O99}), some fraction
may have been formed by a late helium-core flash on the WD cooling
curve.  Because flash mixing would greatly enhance the abundances of
He and C in the envelope, this may affect the pulsations of such
stars.  The calculations of Charpinet et al.\ (1997\markcite{C97})
assume a hydrogen-rich envelope, but it would be interesting to see
what is expected for a flash-mixed envelope.

\subsection{Observational Tests} \label{secobstest}

Although the optical and UV CMDs of NGC~2808 are consistent with a
population of flash-mixed blue-hook stars near the hot end of the
canonical EHB, spectroscopic observations are required to fully
explore the nature of the subluminous stars and validate the
flash-mixing scenario we have described.  Such observations are
ideally suited to STIS, because one would want to observe stars that
occupy various regions in the STIS UV CMD (stars along the BHB, EHB,
and below the EHB).  The center of NGC~2808 is too crowded to observe
well from the ground, but is easily and efficiently available to STIS,
because multiple stars can be placed in its long slits.  Furthermore,
many abundance diagnostics for both the light elements and heavier
metals are available in the UV; thus UV spectroscopy can demonstrate
if carbon and helium are enhanced relative to the other elements.  We
intend to investigate these stars further with STIS spectroscopy.

As explained in \S\ref{secflash}, our calculations cannot accurately
determine if a small amount of residual hydrogen might be left in the
stellar envelope when a star undergoes flash mixing.  This issue can
only be settled by detailed stellar structure calculations that
include the energetics of the proton burning during the flash mixing
phase.  Because of the high gravity in the blue-hook stars,
gravitational settling may bring any residual hydrogen to the surface.
Our calculations show that a thin veneer of hydrogen ($\sim
10^{-6}~M_{\odot}$) at the surface, as indicated by the diffusion
calculations of Fontaine \& Chayer (1997\markcite{FC97}), would not
greatly change the effective temperature of the blue-hook stars: they
would still lie near $\rm T_{eff} \approx 36,000~K$.  However, the
spectrum produced by a star with such a thin hydrogen veneer may not
show the strong abundance enhancements predicted by the flash-mixing
scenario, and instead may look much like a canonical EHB star with a
very small envelope mass.  These stars should still lie below the EHB
in a UV CMD, due to the larger bolometric correction, but they will be
considerably bluer than the flash-mixed blue-hook models shown in
Figure~\fighook.

Evidence for the flash-mixing scenario might also be provided by studying
their post-HB progeny.  As noted in \S\ref{seccmdhb}, the STIS CMD
shows a string of 9 stars that lie hotter and brighter than the
hot end of the HB, in good agreement with theoretical lifetimes that
predict $\sim$9 post-HB stars corresponding to the 46 subluminous EHB
stars.  Figure~\fighook\ shows the post-HB evolution for a flash-mixed
star; the predicted evolution does not reach as bright and hot as
observed for the 9 candidate stars.  However, given the uncertainty in
the surface abundance of the flash-mixed stars (which may change due
to diffusion as the stars evolve toward higher $\rm
T_{eff}$ and lower gravity), we consider at least plausible the
identification of the 9 candidate stars as the AGB-Manqu$\acute{\rm
e}$ progeny of flash-mixed stars.  A spectroscopic search for greatly
enhanced carbon and helium abundances in these 9 hot stars could
provide confirmation.

\section{SUMMARY} \label{secsum}

Our UV CMD of NGC~2808 reveals a significant population of hot stars
directly below the canonical ZAHB.  Like those stars on the canonical
HB, the subluminous stars may be in a core He-burning phase of
evolution, but one that originates in a late helium flash on the WD
cooling curve.  Our evolution models show that such a late flash will
mix the hydrogen envelope with the helium core, which greatly enhances
the envelope helium and carbon abundances.  We have computed new model
atmospheres and synthetic spectra for these flash-mixed stars, which
show that these ``blue-hook'' stars should have lower luminosities and
dramatically different envelope abundances than their counterparts on
the canonical HB.  These abundance enhancements would be detectable in
far-UV spectroscopic observations with HST.

D'Cruz et al.\ (2000\markcite{D00}) invoked blue-hook stars to explain
the subluminous HB stars in their ($F160BW-F555W$, $F555W$) CMD of
$\omega$ Cen.  However, it was difficult to explain the $\sim$0.7~mag
reduction in far-UV luminosity of these stars, because D'Cruz et al.\
(2000\markcite{D00}) assumed that there was no change in the mass or
composition of the hydrogen-rich envelope in their models during the
helium flash; their explanation, based on a small decrease in core
mass, only reduces the bolometric luminosity by $\sim$0.1~mag.
However, their WFPC2/F160BW bandpass overlaps with our STIS/FUV
bandpass, and is sensitive to the same opacity effects discussed here;
if blue-hook stars have enhanced helium and carbon, this would explain
why the subluminous HB stars in $\omega$~Cen extend to such faint
far-UV luminosities below the canonical ZAHB.

Previous observations of NGC~2808 have also reported several gaps in
the HB distribution (Sosin et al.\ 1997\markcite{SDD97}; Bedin et al.\
2000\markcite{BPZ00}).  We have demonstrated that the hottest gap is
likely due to the differences between the canonical EHB stars and the
blue-hook population.  The location of these blue-hook stars, relative
to the canonical EHB, will vary with the bandpasses used for a CMD.
In the STIS UV CMD, the blue-hook stars lie below the canonical EHB
and are not separated by a gap; in optical CMDs, the blue-hook stars
appear as a hot extension of the canonical EHB, separated by a
prominent gap.

Besides $\omega$ Cen and NGC~2808, the only other globular cluster
reported to have a significant EHB population fainter than $M_V =
4.5$~mag is NGC~6273 (Piotto et al.\ 1999\markcite{P99}).  Thus
flash-mixed stars probably do not comprise a significant fraction of
the hot HB stars in classical EHB globular clusters such as NGC~6752
or M13, in which nearly all the EHB stars have $M_V < 4.5$~mag.
However, even non-EHB clusters might have a small number of stars that
experience a late helium-flash.  As noted earlier, Moehler et al.\
(1997\markcite{M97}) report spectroscopy of an EHB star in M15 with
$M_V = 4.7$~mag and a helium abundance of $N_{He}/(N_H + N_{He}) =
0.87$.  This star is an excellent candidate for being a product of a
late helium-core flash.

\acknowledgments Support for this work was provided by NASA through
the STIS GTO team funding.  TMB gratefully acknowledges support at
GSFC by NAS~5-6499D.  AVS gratefully acknowledges support from NASA
Astrophysics Theory Proposal NRA-99-01-ATP-039.  We thank M.A. Wood
for making his white dwarf cooling sequences publicly available.

%%%%%%%%%%%%%%%%%%
%%% REFERENCES %%%
%%%%%%%%%%%%%%%%%%

\vskip 0.5in

%\begin{table}[h]
\noindent
\hskip 0.25in
\parbox{6.5in}{
\begin{center}

{\sc Table \tabcat:} Photometric Catalog
%\caption{Photometric Catalog}

\begin{tabular}{cccccc}
\tableline			  		     		  
RA\tablenotemark{a} & Decl.\tablenotemark{a}& 
$m_{FUV}$ & Error & $m_{NUV}$ & Error\\
(J2000)     & (J2000)    & (mag)   & (mag) & (mag)   & (mag) \\
\tableline			  		     		  
9 12 00.0113 & -64 51 47.334 & 16.43 & 0.02 & 16.92 & 0.01 \\
9 12 00.1639 & -64 51 50.573 & 16.29 & 0.02 & 16.97 & 0.01 \\
9 12 00.1668 & -64 51 53.261 & 16.73 & 0.03 & 17.13 & 0.02 \\
9 12 00.2360 & -64 51 49.498 & 16.41 & 0.01 & 16.94 & 0.01 \\
9 12 00.2451 & -64 51 50.912 & 17.34 & 0.03 & 18.85 & 0.04 \\
\tableline
\end{tabular}
\end{center}

{\small
{\sc Note--} Table~\tabcat\ is available only on-line as a 
machine-readable table.
A portion is shown here for guidance regarding its form and content.
Units of right ascension are hours, minutes, and seconds, and units
of declination are degrees, arcminutes, and arcseconds.\\  

$^{\rm a}$The relative astrometry is very accurate (tenths of 
a 0.025$\arcsec$ STIS pixel), but the absolute astrometry is 
subject to a 1--2$\arcsec$ uncertainty (associated with the 
position of the guide stars).}}
%\end{table}

\end{document}